\documentclass[preprint,superscriptaddress,floatfix,preprintnumbers,nofootinbib,altaffilletter]{revtex4-1}
\pdfoutput=1
\usepackage[utf8]{inputenc}
\usepackage[colorlinks=true,citecolor=blue,linkcolor=blue]{hyperref}
\usepackage{amsmath,amssymb}
\usepackage{epsfig}
\usepackage{graphicx}             
\usepackage{url}
\usepackage{color}
\usepackage{slashed}
\usepackage{multirow}
\usepackage{placeins}
\usepackage[dvipsnames]{xcolor}
\usepackage{epstopdf}

\def\bea{\begin{eqnarray}} 
\def\eea{\end{eqnarray}}

\newcommand{\beq}{\begin{equation}}
\newcommand{\eeq}{\end{equation}}

\newcommand{\GeV}{\text{ GeV}}
\newcommand{\TeV}{\text{ TeV}}

\def\mH{\mathcal{H}}

\def\mL{\mathcal{L}}

\def\tb{\tilde{b}}

\def\ts{\tilde{s}}
\def\tpsi{\tilde{\psi}}

\newcommand{\Eq}[1]{Eq.~(\ref{#1})}

\newcommand{\Ref}[1]{Ref.~\cite{#1}}

\begin{document}

\title{A Light Higgs at  the LHC and the B-Anomalies}


\author{Da Liu}
\affiliation{\mbox{High Energy Physics Division, Argonne National Laboratory, Argonne, IL 60439}}

\author{Jia Liu}
\affiliation{\mbox{Physics Department and Enrico Fermi Institute, University of Chicago, Chicago, IL 60637}}

\author{Carlos E.M. Wagner}
\affiliation{\mbox{High Energy Physics Division, Argonne National Laboratory, Argonne, IL 60439}}
\affiliation{\mbox{Physics Department and Enrico Fermi Institute, University of Chicago, Chicago, IL 60637}}
\affiliation{\mbox{Kavli Institute for Cosmological Physics, University of Chicago, Chicago, IL 60637}}

\author{Xiao-Ping Wang}
\affiliation{\mbox{High Energy Physics Division, Argonne National Laboratory, Argonne, IL 60439}}

\date{\today}
\preprint{EFI-18-6}

\vspace{3.cm}
\begin{abstract}
	After the Higgs discovery, the LHC has been looking for new resonances, decaying into pairs of Standard Model (SM) particles.
	Recently, the CMS experiment observed an excess in the di-photon channel, with a di-photon invariant mass of about 96~GeV.
	This mass range is similar to the one of an excess observed in the search for the associated production of Higgs bosons with 
	the $Z$ neutral gauge boson at LEP, with the Higgs bosons decaying to bottom quark pairs.  On the other hand, the LHCb 
	experiment observed a discrepancy with respect to the SM expectations of the ratio of the decay of $B$-mesons to 
	$K$-mesons and a pair of leptons, $R_{K^{(*)}} = BR(B \to K^{(*)} \mu^+\mu^-)/BR(B\to K^{(*)} e^+e^-)$. This observation provides 
	a hint of the violation of lepton-flavor universality in the charged lepton sector and may be explained by the existence of a 
	vector boson originating form a $U(1)_{L_\mu - L_\tau}$ symmetry and heavy quarks that mix with the left-handed down quarks.
	Since the coupling  to heavy quarks could lead to sizable Higgs di-photon rates in the gluon fusion channel, 
	in this article we propose a common origin of these anomalies identifying a Higgs associated with the breakdown of the $U(1)_{L_\mu - L_\tau}$ 
	symmetry and at the same time responsible to the quark mixing, with the one observed at the LHC. 
	We also discuss the constraints on the identification of the same Higgs with the 
	one associated with the bottom quark pair excess observed at LEP.
\end{abstract}

\maketitle
\tableofcontents

\setcounter{secnumdepth}{2} 
\setcounter{tocdepth}{2}   


\section{Introduction}

The discovery of  a scalar resonance, with properties similar to the ones expected for the Higgs boson in the Standard Model (SM),
has provided evidence for the realization of the simplest Higgs mechanism scenario of electroweak symmetry breaking. The 
couplings of the observed Higgs boson to the SM particles is within a few tens of percent of the ones expected within the SM.  
Small deviations of these coupling with respect to the SM values are still possible, and are expected in extensions of the Higgs sector
that occur in most beyond the SM scenarios. For this reason, since the Higgs discovery, apart from a precise determination of the
Higgs couplings, the LHC has been looking for new scalar resonances, with the di-photon channel being one of the most sensitive ones. 
Recently, the CMS experiment reported a 2.9~$\sigma$ excess in this channel~\cite{CMS:2017yta}, with a di-photon invariant  mass 
of about 95.3~GeV. This excess was mildly present in the 8~TeV run~\cite{CMS-PAS-HIG-14-037}, but became prominent only in 
the 13~TeV run.  While the ATLAS experiment did not  observe  any significant excess in this mass region in the 8 TeV 
run~\cite{,Aad:2014ioa},  it has not yet reported the results of a similar search in the 13~TeV run.

Searches for Higgs boson resonances produced in association with the $Z$ gauge boson, with Higgs bosons decaying into bottom-quark
pairs, were conducted at LEP.  The combination of the results of the four experiments, ALEPH, DELPHI, L3 and OPAL, led to the presence
of a 2.3~$\sigma$ local excess at an invariant mass of about 95--100~GeV~\cite{Barate:2003sz}. The agreement between the invariant mass 
of the excesses observed at LEP and CMS calls for a possible common origin of these two signatures~\cite{Cao:2016uwt, 
Mariotti:2017vtv, Crivellin:2017upt, Fox:2017uwr, Haisch:2017gql, Vega:2018ddp}. 

On the other hand, the LHCb experiment has reported an intriguing hint of the violation of lepton-flavor universality in the decay of
$B$-mesons into $K$-mesons and lepton pairs~\cite{Aaij:2014ora,Aaij:2017vbb}, namely
\begin{equation}
R_K = \frac{BR(B \to K \mu^+\mu^-)}{BR(B\to K e^+e^-)} = 0.745^{+0.097}_{-0.082}
\end{equation}
while
\begin{equation}
R_{K^*} = \frac{BR(B \to K^* \mu^+\mu^-)
}{BR(B\to K^* e^+e^-)}  = 0.660^{+0.113}_{-0.074}.
\end{equation}

\begin{figure}
	\centering
	\includegraphics[width=0.5 \textwidth]{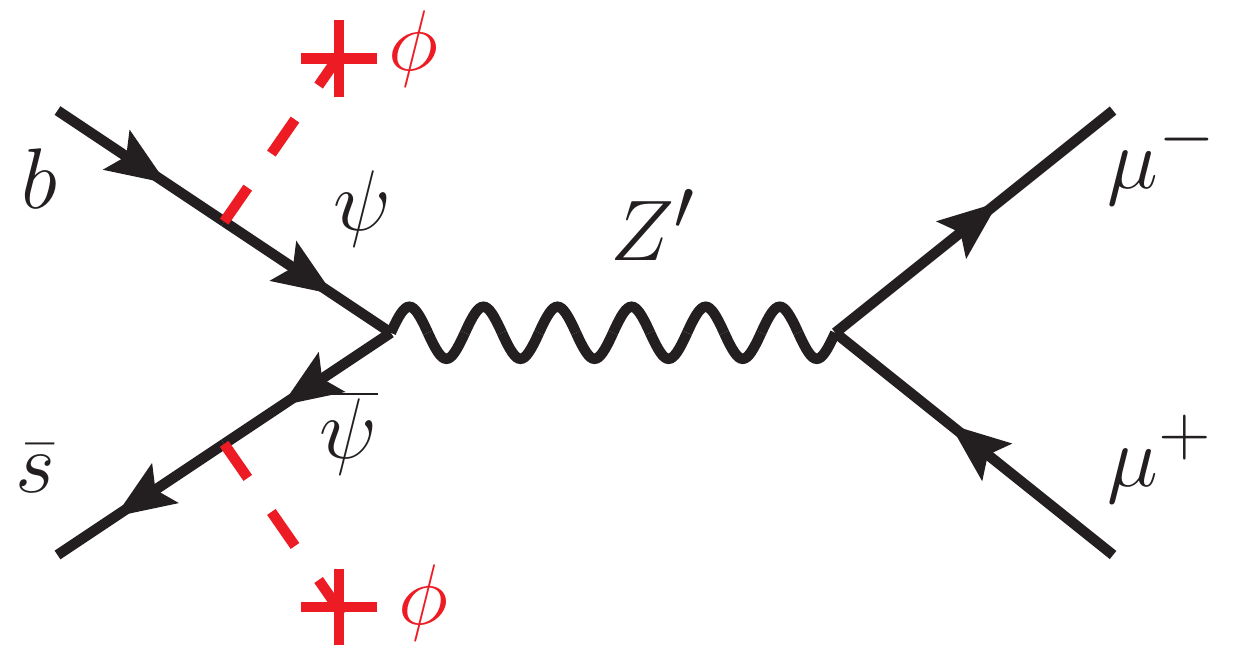} 
	\caption{  Feynman diagram showing how the Higgs $\phi$ induces an off-diagonal coupling of the new gauge boson $Z'$ to 
	            the strange and bottom quarks via the  mixing of the SM quarks with the heavy vector-like quark $\psi$.  This process induces  a 
		contribution to the $C_9^{NP}$ effective operator which is responsible for the B-anomalies. A sizable coupling of the 
		Higgs $\phi$ to di-photons and di-gluons will be induced via the coupling to the heavy quark, and hence $\phi$ could be
	potentially identified with the one observed by the CMS~\cite{CMS-PAS-HIG-14-037, CMS:2017yta} experiment in the di-photon channel. } 
	\label{fig:diagram}
\end{figure}

A possible explanation of these anomalies may be provided by the introduction of a gauge boson associated with
the $U(1)_{L_\mu - L_\tau}$ symmetry~\cite{Altmannshofer:2014cfa,Crivellin:2015mga,Altmannshofer:2015mqa,Altmannshofer:2016oaq,Altmannshofer:2016jzy,Alonso:2017uky,Bonilla:2017lsq,Nomura:2018vfz,Chen:2017usq,Ko:2017yrd} or other flavor symmetry~\cite{Crivellin:2015lwa,Celis:2015ara,Falkowski:2015zwa, Alok:2017sui}.  
The absence of a coupling to 
electrons explains the deviation of the above ratios with respect to one, the value expected within the SM. 
In order to allow the coupling of the new gauge boson to the $B$
meson, one can introduce new vector like quarks that mix with the ordinary third and second generation quarks 
and that is charged under the new gauge symmetry, see the illustrative diagram in Fig. \ref{fig:diagram}. 
Such a quark mixing may be induced by the vacuum 
expectation value of a new Higgs boson that is neutral under the SM gauge symmetries. The coupling of this Higgs boson to the vector
like quarks can induce sizable couplings to photons and gluons, which could be searched in the di-photon channel. 
Similarly, a mixing of such a Higgs with the SM-like Higgs boson can induce a coupling to the SM gauge bosons, 
while couplings to quarks are induced by Higgs and quark mixing effects. Hence, such a Higgs can also be produced in 
association with gauge bosons, subsequently decaying to bottom quark pairs. 
Therefore, a very natural question is whether such Higgs boson can be identified with the one that is observed by the CMS and 
LEP experiments. We analyze the signal and existing constraints in this paper and answer this interesting question
in the conclusion section.

In this article, we shall describe the simplest scenario that can lead to a realization of this idea. In section~\ref{sec:TheModel}
we shall present the model and describe the interaction of the Higgs and gauge bosons. In section~\ref{sec:LightHiggs}
we shall discuss the possibility that a Higgs boson associated with the breakdown of the new $U(1)_{L_\mu - L_\tau}$
symmetry leads to an explanation of the CMS excess and explain the constraints associated with a simultaneous
explanation of the LEP excess. In section~\ref{sec:BAnomalies} we shall discuss the explanation of the anomalies
seen in B-meson decays.  In section~\ref{sec:Constraints} we shall concentrate on the collider and flavor physics
constraints on this model. Finally, we reserve section~\ref{sec:Conclusions} for our conclusions, and the appendices
for technical details associated with the scalar mixing parameters and the obtention of the proper CKM matrix
elements within this model.

\section{A model with an extra $U(1)$, light Higgs and  heavy vector-like quarks}
\label{sec:TheModel}

In this section, we shall introduce a gauge extension of the Standard Model (SM) to include two
extra light Higgs $S$, $\phi$ and one extra gauge boson $Z'$ associated with the $U(1)_{L_\mu- L_\tau}$
symmetry \cite{PhysRevD.44.2118,Baek:2001kca,Foot:1994vd,Salvioni:2009jp,Heeck:2011wj,PhysRevD.85.115016,
Harigaya:2013twa,Ma:2001md,Ibe:2016dir,Biswas:2016yan,Elahi:2015vzh,Gninenko:2018tlp,Cao:2017sju,Biswas:2017ait,
Baek:2017sew,Chen:2017usq,Elahi:2017ppe,Asai:2017ryy,Kaneta:2016uyt,Biswas:2016yjr, Tang:2017gkz,
Ko:2017quv, Arcadi:2018tly,Kamada:2018zxi,Biswas:2018yus}. 
The gauge charge $L_\mu- L_\tau$ is not flavor universal, but diagonal. We also need one 
extra vector-like heavy quark $\psi_{L,R}$ charged under both SM gauge and $U(1)_{L_\mu- L_\tau}$.
The model is minimal to address the light Higgs and B-anomalies. The SM and new physics (NP) particles are listed in 
Table \ref{tab:gaugecharge} along with their gauge charges.
$q_L^i$ ($u_R^i, d_R^i$) are the $i$th generation left-handed (right-handed) SM quarks, where
$i = 1,2 ,3 $ is the SM quark generation index. $\ell^{e,\mu,\tau}$ and $e_R, \mu_R, \tau_R$ are three generations 
of SM left-handed and right-handed leptons. $S$ and $\phi$ are NP Higgs bosons, which are singlet under the
SM gauge symmetry and are responsible to give mass to $U(1)_{L_\mu - L_\tau}$ gauge boson $Z'$. The fields $\psi_{L,R}$ the
chiral components of a vector-like heavy quark, which has the same SM charge as $q_L^i$ and also charged under 
$U(1)_{L_\mu - L_\tau}$. The charge assignments 
of $\phi$ and $\psi$ under the $U(1)_{L_\mu - L_\tau}$ symmetry need to satisfy the relation $Q_{\mu - \tau}^\phi = \pm Q_{\mu - \tau}^\psi$ 
to allow for the appropriate Yukawa couplings.
Without loss of generality, it can be taken to be $Q_{\mu - \tau}^\phi = - Q_{\mu - \tau}^\psi = 1/2$. The anomaly-free condition 
is satisfied because $U(1)_{L_\mu - L_\tau}$ is vector-like in the NP particle content. It is also anomaly-free within the SM sector. 
The vacuum expectation value of the field $\phi$ may induce a
mixing between the SM quarks and the NP heavy quarks. 
In the following sections, we will discuss the model in details in  the Higgs, gauge boson and quark sectors.

	\begin{center}
\begin{table}[htb]
\scalebox{1.1}{%
		\begin{tabular}{|c||c|c|c||c|}
			\hline 
			filed & $SU(3)_C$ &$SU(2)_L$  &$U(1)_Y$ & $U(1)_{L_\mu - L_\tau}$ \\  \hline
			\hline
			$H$ & 1 & 2   &  $\frac{1}{2}$ & 0 \\
			\hline
			$q_L^i$   &  3  & 2  & $\frac{1}{6}$    & 0  \\
			\hline
			$u_R^i$   &   3   & 1  & $\frac{2}{3}$ & 0   \\  
			\hline
			$d_R^i$  & 3  & 1  &  $-\frac{1}{3}$  & 0  \\
			\hline	\hline
			$\ell^e$    &   1  &2   & $-\frac12$ &0 \\  
			\hline
			$\ell^\mu$    &   1  &2   &$ -\frac12 $&1 \\  
			\hline
			$\ell^\tau$    &   1  &2   &$ -\frac12$ &-1\\  
			\hline
			$e_R$    &   1  &1   & -1 &0 \\  
			\hline
			$\mu_R$    &   1  &1   & -1 &1 \\       
			\hline
			$\tau_R$    &   1  &1   & -1 &-1\\         
			\hline	\hline
			$S$  & 1   & 1  & 0  & 1   \\ 
			\hline
			$\phi$  & 1   & 1  & 0  & $\frac{1}{2}$   \\     		    
			\hline
			$\psi_{L,R}$   &  3  & 2   & $\frac{1}{6}$    & $ -\frac{1}{2}$  \\
	                 \hline
		\end{tabular}}
		\caption{
		The particle content and their $SU(3)_C\times SU(2)_L \times U(1)_Y \times U(1)_{L_\mu - L_\tau}$ gauge charge.
		$i = 1,2 ,3 $ is the SM quark generation index. $S$ and $\phi$ are the NP Higgs, and $\psi_{L,R}$ are vector-like heavy quarks. 
		The charge assignment $Q_{\mu - \tau}^\phi = -Q_{\mu - \tau}^\psi$  allows for the appropriate 
		Yukawa couplings. Without loss of generality, it is taken to be $1/2$. }
	\label{tab:gaugecharge}
\end{table}
\end{center}

\subsection{The Higgs sector and gauge boson sector}

In the Higgs sector of this model, we have two extra Higgs $S$, $\phi$. We can write down the Lagrangian for the Higgs bosons,
\begin{align}
\mathcal{L}_{\rm Higgs} &= \left(D_\mu \phi \right)^\dagger D^\mu \phi  +\left(D_\mu S \right)^\dagger D^\mu S \nonumber \\
&- \mu_\phi^2 \phi^{*} \phi 
- \lambda_\phi \left( \phi^{*} \phi  \right)^2  - \mu_S^2 S^{*} S 
- \lambda_S \left( S^{*} S \right)^2 - \mu^2 H^{\dagger} H - \lambda \left( H^\dagger H \right)^2 \nonumber \\
&- \lambda_{\phi h} \left( \phi^{*} \phi  \right) \left( H^\dagger H  \right)  - \lambda_{S h} \left(S^{*} S \right) \left( H^\dagger H  \right) - \lambda_{S \phi} \left(S^{*} S \right) \left( \phi^{*} \phi  \right).
\label{eq:HiggsLagragian}
\end{align}
The covariant derivative $D_\mu = \partial_\mu - i g_D Z'_\mu Q_{L_\mu - L_\tau}$, with $g_D$ the gauge coupling of
$U(1)_{L_\mu - L_\tau}$. The first line in $\mathcal{L}_{\rm Higgs}$ are kinetic terms for $\phi$ and $S$, which give mass to $Z'$ 
after symmetry breaking,  and others are scalar potential terms. The couplings $\lambda_{\phi h}$, $\lambda_{S h}$ and $\lambda_{S\phi }$ 
define the  Higgs portal terms, which induce the mixing between scalars. The scalar $\phi$, $S$
and SM Higgs $H$ get vevs and can be expanded near the minimum as 
$\phi = \frac{1}{\sqrt{2}} \left(\phi^0 + v_D\right)$, $S = \frac{1}{\sqrt{2}} \left(s^0 + v_S\right)$ 
and $H = \left( 0, \frac{1}{\sqrt{2}} \left(h^0 + v\right)  \right)$. 

The gauge boson $Z'$ has no mass mixing with the SM gauge bosons. Only the two kinetic terms in the first line of 
Eq. (\ref{eq:HiggsLagragian}) give the $Z'$ mass, which reads
\begin{align}
m_{Z'} = g_D \sqrt{\frac{1}{4}  v_D^2+ v_S^2}.
\end{align}
In our analysis, we assume the observed 96~GeV di-photon excess observed at CMS is predominantly associated with the CP-even component of
$\phi^0$. This means $v_D$ is nearly the same order of the SM Higgs vev $v$. If there is no another Higgs $S$, the $Z'$ mass is 
too light and severely constrained by di-jet and di-lepton constraints in Section \ref{sec:Zpconstraint}. We therefore introduce
another scalar $S$ with unit charge under $U(1)_{L_\mu - L_\tau}$,  and that contribute to the $Z'$ mass, as described above. 
Therefore, the $Z'$ mass can be as heavy as several TeV with a large value of $v_S$ and we treat it as a free parameter thereafter. 
In the later analysis, we could find out $m_{Z'} \sim 4.1$ TeV, thus $m_{s^0}$ should be 
of the same order.  We shall not consider the phenomenology of such a heavy Higgs, as it could easily decouple from the SM Higgs
and $\phi$ phenomenology. In our analysis, we assume $\lambda_{Sh}$ and $\lambda_{S\phi}$ small, so that complemented with the heaviness 
of  $s^0$ we can neglect its mixing with the light Higgs bosons and  only consider the mixing between H and $\phi$.
On the other hand, $S$ has different $U(1)_{L_\mu - L_\tau}$ charge to $\Psi_{L,R}$ and SM quarks. It will not affect the quark sector mixing 
and we do not need to consider it any more.

For the scalar mixing, after ignore the effect of $S$, we only consider two scalars $\phi^0$ and $h^0$ with five free parameters 
$\mu_\phi, \mu, \lambda_\phi, \lambda$ and $\lambda_{\phi h}$. The five parameters can be traded to five physical observables, two 
vevs $v_D$ and $v$, two masses $m_{\tilde{h}}$ and $m_{\tilde{\phi}}$, and one mixing angle $\sin \alpha$. $\tilde{h}$ and 
$\tilde{\phi}$ are the mass eigenstates and related to $\phi^0$ and $h^0$ by 
\begin{align}
\left(\begin{array}{c}  h^0 \\ \phi^0 \end{array}\right) =  \left(\begin{array}{cc}
\cos\alpha & -\sin\alpha \\
\sin\alpha  & \cos\alpha \end{array}
\right)\left(\begin{array}{c} \tilde h\\ \tilde \phi \end{array}\right) .
\end{align}
The five physical observables are the more useful model parameters. The relations with the old parameters
are given in Appendix \ref{sec:scalarmixing}.

\subsection{ The quark sector }

After the scalar sector, we are going to specify the mixings between the SM quarks and heavy
vector-like quarks, which are responsible to induce the flavor violating couplings needed for 
the explanation of the B-anomalies. 
The most general interactions in the quark sector are given by:
\begin{align}
\mathcal{L}_q & = i \bar{q}^i_L \slashed D q_L^i +  i \bar{u}^i_R \slashed D u_R^i +  i \bar{d}^i_R \slashed D d_R^i  +  i \bar{\psi} \slashed D \psi - m_{\psi}  \bar{\psi} \psi  \nonumber  \\ 
& - \left( \bar{q}^i_L y^{ij}_{u} \tilde H u_R^j+ \bar{q}_L^i y_{d}^{ij} H d_R^{j} + H.c.\right)  
- \left(  \sqrt{2}\lambda_i \bar{q}_L^i \phi \psi_R +H.c. \right)  \,, 
\end{align}
where $q_L$ are left-handed SM quarks, $u_R$ and $d_R$ are up-type and down-type right-handed SM quarks. 
$\psi$ is the heavy vector-like quark, $H$ is the SM Higgs and $\phi$ is the NP Higgs. $i, j$ are the generation
index, $y_u, y_d$ are the SM Yukawa couplings. $\tilde{H}$ is defined as $\tilde{H} \equiv
i \sigma_2 H^* $. The $SU(2)_L$ doublet $\psi$ can be written as components $(\psi^u, \psi^d)$, which are mass degenerate. 
After the scalars obtained vevs, the mass and interaction terms for quarks are,
\begin{align}
\mathcal{L}_{Higgs} =  - \sum_{q=u, d} \left(\bar{q}^3_{L}, \bar{q}^2_{L}, \bar{\psi}^q_{L} \right) \cdot 
\left(M_q + \Lambda_q \right) \cdot \left(q^3_{R}, q^2_{R}, \psi^q_{R} \right)^T  +H.c. \,,
\label{eq:fermionmassHiggs}
\end{align}
where the sum over $q = u, d$ goes over the components of $SU(2)_L$ doublet. 
Since SM quarks are not charged under $U(1)_{L_\mu - L_\tau}$, their Yukawa couplings are SM-like
with 9 free parameters. For simplicity, we include only the third and second generation of SM quarks together
with the heavy quark $\psi$ in Eq.~(\ref{eq:fermionmassHiggs}), to illustrate the essence of NP phenomenology.
To further simplify the calculation, SM Yukawa couplings are chosen to be flavor diagonal.  We shall relegate
to the Appendix the discussion of the obtention of the CKM matrix elements, which is not relevant for our
analysis. Thus, the mass matrix $M_q$ and interaction matrix $\Lambda_q$ are 
\begin{align}
M_q=\left(\begin{array}{ccc}
m_{q^3} & 0 & \lambda_3 v_D \\
0 & m_{q^2} &\lambda_2 v_D\\
0 & 0  &m_\psi
\end{array}\right) , \quad
\Lambda_q = \left(\begin{array}{ccc}
\frac{m_{q^3}}{v} h^0 & 0 & \lambda_3 \phi^0 \\
0 & \frac{m_{q^2}}{v} h^0 &\lambda_2 \phi^0\\
0 & 0  & 0
\end{array}\right) .
\end{align}
Note for the up-type quarks $u^3 = t$ and $u^2 = c$ and for down-quark $d^3 = b$ and $d^2 = s$. In the matrices
$M_q$ and $\Lambda_q$,  only the diagonal mass elements are modified when changing from $u$ to $d$, while the
terms proportional to $\lambda_{2,3}$ stay the same. Since $\psi$ is heavy vector-like quark, the masses have the sequence 
$m_\psi \gtrsim m_{q^3} > m_{q^2}$, where as we will show $\lambda_2 v_D > m_\psi $ is necessary to obtain a large enough
mixing between heavy quarks and SM 2nd generation quarks. For the up sector,  since the top quark mass is of the
order of the weak scale, $m_{q^3}$ can be comparable with $m_\psi$.  
To avoid large mixing between the top and the  3rd generation, and hence large Yukawas to obtain the proper top mass, 
we further assume $\lambda_2 \gg \lambda_3$ which suggests the heavy quark mostly mixes with 2nd generation 
left-handed SM quarks. After diagonalizing $M_q$, we get the mass of the quark mass-eigenstates,
\begin{align}
m_{\tilde{q}^3}^2 & \approx m_{q^3}^2 -\frac{m_{q^3}^2 \lambda_3^2v_D^2}{\lambda_2^2v_D^2+m_\psi^2-m^2_{q^3}} 
= m_{q^3}^2(1-\tan^2\theta_3)   , \\
m_{\tilde{q}^2}^2 & \approx m_{q^2}^2\frac{m_\psi^2}{m_\psi^2+\lambda_2^2v_D^2}= m_{q^2}^2\cos^2\theta_2 , \\
m_{\tilde \psi}^2 & \approx m_\psi^2 + v_D^2\lambda_2^2+\frac{\lambda_3^2v_D^2\left(\lambda_2^2v_D^2+m_\psi^2\right)}{\lambda_2^2v_D^2+m_\psi^2 -m^2_{q^3}} +m_{q^2}^2 \frac{\lambda_2^2 v_D^2}{\lambda_2^2v_D^2+m_\psi^2} 
\simeq m_\psi^2 + \lambda_2^2 v_D^2 , \label{eq:heavyQmass}
\end{align}
where in the last equality we only keep the leading term and we define $\tan\theta_2 \equiv \frac{\lambda_2v_D}{m_\psi}$ 
and $\tan\theta_3  \equiv  \frac{\lambda_3 v_D}{\sqrt{\lambda_2^2v_D^2 +m_{\psi}^2 -m^2_{q^3}}}$. 
Following the mass and coupling assumption, $\tan\theta_3 \ll 1$. As explained above,  
to fit the B-meson decay anomalies, $\lambda_2 v_D $ needs to be larger than $m_\psi$, which suggests $\sin \theta_2 $ 
and $\cos \theta_2 $ are of $\mathcal{O}(1)$.
Note that the mass of $\tilde{\psi}_u$ and $\tilde{\psi}_d$ are almost degenerate, only different by small 2nd 
generation quark mass $m_{q^2}$ and coupling $\lambda_3$.

The mixing matrices $U^q_{L,R}$ connect flavor and mass eigenstates as
\begin{align}
\left(q^3, q^2, \psi^q \right)^T_{L,R} = U^q_{L,R} \cdot \left(\tilde{q}^3, \tilde{q}^2, \tilde{\psi}^q \right)^T_{L,R} ,
\end{align}
where the states with tilde are the mass eigenstates for the quarks. The mixing matrices at leading order are

\begin{align}
U^q_L& \approx \left(
\begin{array}{ccc}
1+\frac{1}{2} t^2_{\theta_3} &0 & t_{\theta_3} \\
- s_{\theta_2} t_{\theta_3} & c_{\theta_2}\left( -1 + \frac{m^2_{\tilde {q}^2}}{m^2_{\tilde \psi}}t^2_{\theta_2} \right)& 
s_{\theta_2}  \left(1+\frac{m^2_{\tilde {q}^2}}{m^2_{\tilde \psi}} - \frac{1}{2} t^2_{\theta_3} \right)\\
-c_{\theta_2} t_{\theta_3} & s_{\theta_2} \left(1 + \frac{m^2_{\tilde {q}^2}}{m^2_{\tilde \psi}}  \right) &
c_{\theta_2} \left(1-\frac{m^2_{\tilde {q}^2}}{m^2_{\tilde \psi}} t^2_{\theta_2} - \frac{1}{2} t^2_{\theta_3} \right)
\end{array}
\right)   , \label{eq:UL}\\
U^q_R&  \approx \left(
\begin{array}{ccc}
1 & \frac{m_{\tilde {q}^2}}{m_{\tilde {q}^3}} t_{\theta_2} t_{\theta_3} &\frac{m_{q^3}}{m_{\tilde \psi}} t_{\theta_3} \\
- \frac{m_{\tilde {q}^2}}{m_{\tilde {q}^3}}  t_{\theta_2} t_{\theta_3} & 1& \frac{m_{q^2}}{m_{\tilde \psi}} s_{\theta_2} \\
-\frac{m_{q^3}}{m_{\tilde \psi}} t_{\theta_3} & -\frac{m_{q^2}}{m_{\tilde \psi}} s_{\theta_2} &1
\end{array}
\right) , \label{eq:UR}
\end{align}
where $s_{\theta_2}, c_{\theta_2}, t_{\theta_2}$ are abbreviations for $\sin\theta_2$, $\cos\theta_2$
and $\tan\theta_2$.

We see that $U_R^q$ is close to identity matrix with off-diagonal terms suppressed by small SM quark 
mass $m_{q^2,q^3}$ or $\tan\theta_3$. Because the  heavy vector-like quark has the same gauge SM charge as 
the left-handed
SM quarks, thus the mixing dominantly happens between the left-handed quarks. For the left-handed
mixing $U_L^q$ in Eq.~(\ref{eq:UL}), the leading term are the same for up-type an down-type quarks, 
if neglecting the SM quark masses.

The SM Cabibbo-Kobayashi-Maskawa (CKM) matrix elements should be obtained from  
$V_{\text{CKM}}  \equiv U^{u\dagger}_L U^d_L$.  However, since we did not include off-diagonal matrix elements
for the SM-quarks, all non-vanishing elements will be proportional to rations of the SM-quark masses and the
heavy fermion masses. In reality, the
successful generation of CKM matrix needs to include the 1st generation SM quark mass and the off-diagonal
terms in SM Yukawa couplings. In principle, with 9 free mass parameters in SM quark sector, there should
be no problem to generate CKM matrix. 
We will demonstrate the generation of all the $V_{\text{CKM}}$ terms in the Appendix \ref{sec:CKM}, 
and show that it does not affect the NP phenomenology we are interested here. 

After the discussion
of the quark sector, it is worth to mention that the charged leptons can easily get mass with diagonal SM Yukawa 
terms. For neutrino mass and Pontecorvo–Maki–Nakagawa–Sakata (PMNS) matrix under gauge interaction 
$U(1)_{L_\mu- L_\tau}$,  it is generally easy to accommodate. The reason is that $L_\mu- L_\tau$ gauge satisfies
$\mu \leftrightarrow \tau$ permutation which leads to mixing angle $\theta_{23}$ close to maximal ($45^\circ$) 
while having a vanishing mixing angle $\theta_{13}$ at leading order (see review \cite{Xing:2015fdg}).
Since neutrino mass is not relevant of the NP phenomenology we are interested in, we do not modify the lepton
sector further.  

\subsection{Light and heavy quark interactions}
\label{sec:SVquarks}

In this subsection, the flavor eigenstate scalars and quarks are rotated into mass eigenstates. In the mass 
eigenstate, the Yukawa interactions for scalars and quarks are,
\begin{align}
\mathcal{L}_{Higgs} & \supset \frac{m_{\tilde b}}{v}\left(\frac{v}{v_D} t^2_{\theta_3} 
\left(\frac{m^2_{\tilde b}}{m^2_{\tilde \psi}   }  -c^2_{\theta_2}\right) s_\alpha+\left(1-\frac{t^2_{\theta_3}}{2}\right)c_\alpha\right) 
\tilde{h} \bar{\tilde b}\tilde{b}  \nonumber \\
& +  \frac{m_{\tilde b}}{v}\left(\frac{v}{v_D} t^2_{\theta_3} \left(\frac{m^2_{\tilde b}}{m^2_{\tilde \psi}} -c^2_{\theta_2}\right) 
 c_\alpha -\left( 1-\frac{t^2_{\theta_3}}{2}\right)s_\alpha\right)  
\tilde{\phi} \bar{\tilde b}\tilde{b} \nonumber \\
&+ \frac{m_{\tilde s}}{v } \left(\frac{v}{v_D} s^2_{\theta_2}  s_\alpha -  c_\alpha \right) 
\tilde{h} \bar{\tilde s}\tilde{s}
+ \frac{m_{\tilde s}}{v}  \left(\frac{v}{v_D} s^2_{\theta_2}  c_\alpha +  s_\alpha \right)  \tilde{\phi}\bar{\tilde s}\tilde {s}   \nonumber \\ 
&+ \frac{m_{\tilde \psi}}{v}\frac{v}{v_D} s_\alpha\left(s^2_{\theta_2}+\frac{t^2_{\theta_3}}{2}\left(1+c^2_{\theta_2}\right)\right)
\tilde{h}\bar{\tilde \psi}\tilde{\psi}
+ \frac{m_{\tilde \psi}}{v}\frac{v}{v_D} c_\alpha \left(s^2_{\theta_2} +\frac{t^2_{\theta_3}}{2}\left(1+c^2_{\theta_2}\right)\right)  
\tilde \phi \bar{\tilde \psi}\tilde{\psi} \nonumber  \\ 
&+ \left( b\to t, s\to c \right) + \ldots \,,
\label{eq:scalarQuarks}
\end{align}
where the diagonal interactions with scalars are given in Eq.~(\ref{eq:scalarQuarks}). The dots in the last line are for the omitted off-diagonal
interactions, which dominantly opens the decay of $\tilde{\psi}$ into 2nd generation SM quarks.  
The decay width of process $\tilde{\psi}^q \to \tilde{\phi} q^2$ is proportional to 
$c_\alpha^2 s^2_{\theta_2} \left( m_\psi/v_D\right)^2$, while the other decays $\tilde{\psi}^q \to \tilde{\phi} q^3$ or
$\tilde{\psi}^q \to \tilde{h} q^2$ are suppressed by either $t^2_{\theta_3}$ or $s^2_\alpha$.
The off-diagonal interactions will not contribute to scalar $\tilde{h}, \tilde{\phi}$ couplings to gluon-gluon
and photon-photon, thus they are irrelevant for the NP phenomenology.

The gauge interactions between $Z^\prime$ and quark mass eigenstates at leading order are given below:
\begin{align}
\label{eq:ZpFermions}
\mathcal{L}_{Z'}^{\text{quark}}&=- \frac{g_D}{2} t^2_{\theta_3} c^2_{\theta_2} Z'_\mu \bar{\tilde b}_L\gamma^\mu \tilde b_L  
- \frac{g_D}{2}Z'_\mu  s^2_{\theta_2}\bar{\tilde s}_L\gamma^\mu \tilde s_L   
\nonumber \\
 & +\frac{g_D}{2}  Z'_\mu \left[ \left( t^2_{\theta_3} -1 \right)c^2_{\theta_2}  \bar{\tilde \psi}^d_L\gamma^\mu \tilde \psi_L^d 
+  \left( t^2_{\theta_3} -1 \right)\bar{\tilde \psi}_R^d \gamma^\mu \tilde \psi_R^d \right] \nonumber \\ 
& + \frac{g_D}{2} Z'_\mu s_{\theta_2}c_{\theta_2} t_{\theta_3} \bar{\tilde b}_L\gamma^\mu \tilde s_L+ H.c.  \nonumber \\
&+  \frac{g_D}{2} Z'_\mu \left[-c^2_{\theta_2}  t_{\theta_3} \bar{\tilde b}_L\gamma^\mu \tilde \psi_L^d 
- \frac{m_{\tilde b} }{m_{\tilde \psi}} t_{\theta_3}\bar{\tilde b}_R\gamma^\mu \tilde \psi_R^d \right]  + H.c.  \nonumber \\
&+ \frac{g_D}{2} Z'_\mu \left[ s_{\theta_2}c_{\theta_2} \left( \frac{t^2_{\theta_3}}{2} -1\right)\bar{\tilde s}_L\gamma^\mu \tilde \psi_L^d 
+ \frac{m_{s}}{m_{\tilde \psi}} s_{\theta_2}\bar{\tilde s}_R\gamma^\mu \tilde \psi_R^d \right]  + H.c. \nonumber \\
& + \left( b\to t, s\to c \right)   ,
\end{align}
where the first two lines are diagonal interactions, and the lines from three to five are off-diagonal 
interactions with quarks.  Such interactions induce $Z'$ decay into $\bar{\tilde \psi} + \tilde{b},\tilde{s} $ for our benchmark point $m_{Z'}> m_{\tilde \psi}$.
The interaction with up-type quarks are similar, simply by substitution $b\to t$ and $s\to c$. 
Note the flavor violating interaction in line three is required to solve the B-anomalies. However, it also induces $B-\bar{B}$ mixing and
$b \to s \gamma$ via 1-loop, which constraints are investigated in Section~\ref{sec:flavorconstraint}. 

\begin{figure}
	\centering
	\includegraphics[width=0.5 \textwidth]{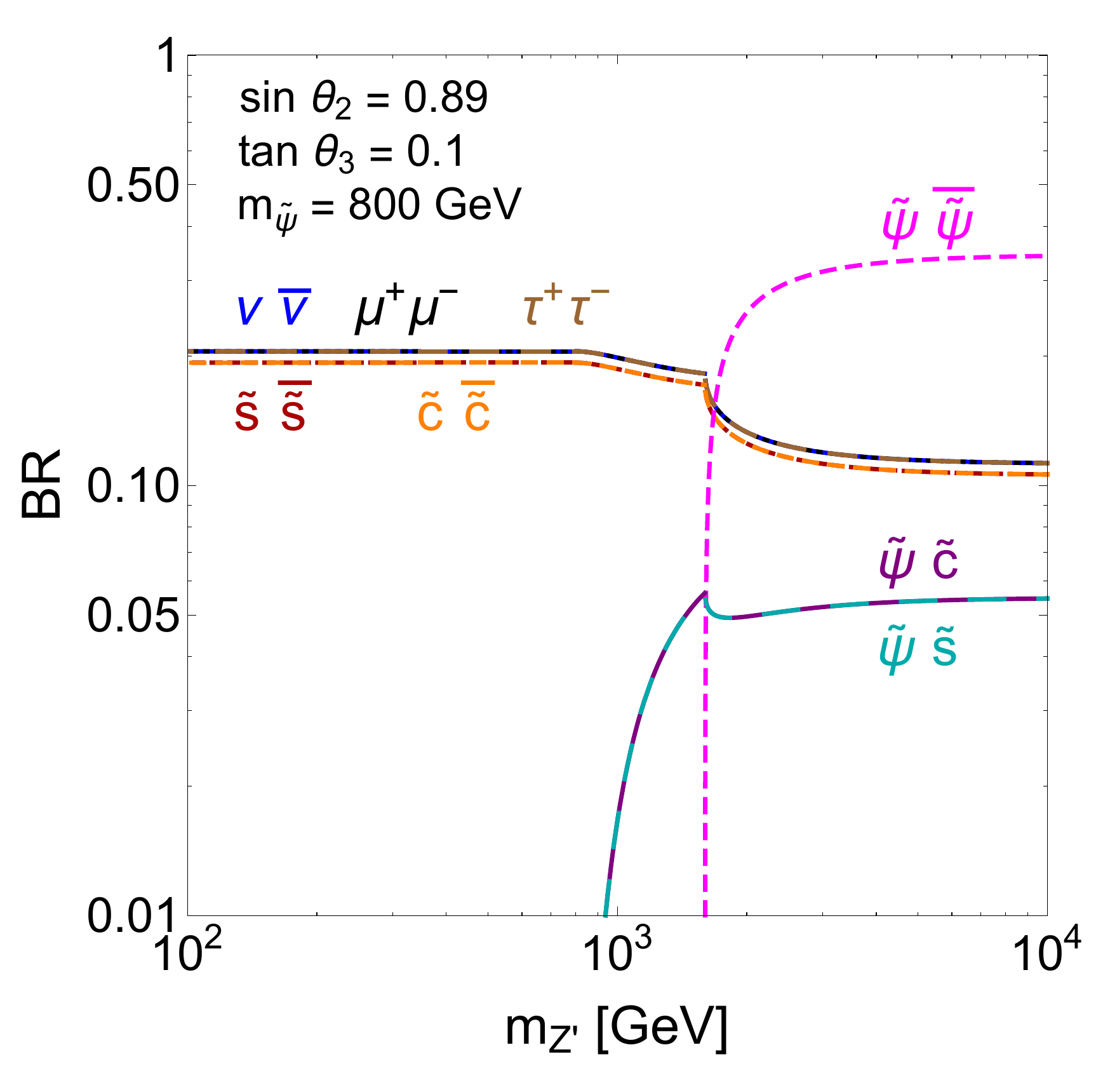} 
	\caption{ The $Z'$ decay branching ratios. The label ``$\bar{\nu} \nu$" includes both 
		$\bar{\nu}_\mu \nu_{\mu }$ and $\bar{\nu}_\tau \nu_{\tau }$. The label ``$\tilde \psi \tilde {c} (\tilde s) $" include both ``$\bar{\tilde \psi} \tilde {c} (\tilde s)$" and ``$\tilde \psi \bar{\tilde {c}} (\bar{\tilde s})$".}
	\label{fig:Zp-BR}
\end{figure}

The lepton interactions of $Z'$ are 
\begin{align}
\mathcal{L}_{Z'}^{\text{lepton}} & =  g_D Z^\prime_\mu \left[\bar {\mu} \gamma^\mu \mu-\bar {\tau} \gamma^\mu \tau + \bar {\nu}_{\mu L} \gamma^\mu \nu_{\mu L} -\bar {\nu}_{\tau L} \gamma^\mu \nu_{\tau L} \right] + H.c.  ,
\label{eq:Zprimelepton}
\end{align}
Due to the charge assignments, in the absence of mixing, the $Z^\prime$ coupling to leptons would be  2 times larger 
than the one to heavy quarks. 
We calculate the decay branching ratios (BR) for $Z'$ into heavy vector-like quark $\bar{\tilde{\psi}} \tilde{\psi}$, 
2nd generation SM quarks $\bar{\tilde{s}} \tilde{s}$ and $\bar{\tilde{c}} \tilde{c}$, 2nd and 3rd generation
SM charged leptons $\mu^+ \mu^-$, $\tau^+\tau^-$  and neutrinos
$\bar{\nu}_\mu \nu_{\mu }$, $\bar{\nu}_\tau \nu_{\tau }$. The width of the decay to 3rd generation quarks are 
suppressed by $t_{\theta_3}^4$. The only significant flavor off-diagonal decays for $Z'$ are
to 2nd generation quark and heavy vector-like quarks $\bar{\tilde{s}} \tilde{\psi}^d, \bar{\tilde{\psi}}^d \tilde{s}$
and $\bar{\tilde{c}} \tilde{\psi}^u, \bar{\tilde{\psi}}^u \tilde{c}$. The flavor off-diagonal decays for $Z'$ to 3rd
generation are suppressed by $t_{\theta_3}^2$. The decay BRs for $Z'$ are given in Fig.~\ref{fig:Zp-BR}.  

Before closing this subsection, we calculate the leading order couplings between the SM $Z$ boson and SM 
quarks, which are modified due to mixing effects,
\begin{align}
\mathcal{L}_{Z}& = \frac{g }{c_w} \left( -\frac{1}{2} + \frac{1}{3} s_w^2 \right) Z_\mu \left[\bar{\tilde b}_L\gamma^\mu \tilde b_L + \bar{\tilde s}_L\gamma^\mu \tilde s_L + \bar{\tilde \psi}_{L}^d \gamma^\mu \tilde \psi_L^d \right]  \nonumber \\ 
&+ gZ_\mu \left[\left(\frac{ s_w^2}{3c_w}-\frac{ m^2_{\tilde b} t^2_{\theta_3}}{2m^2_{\tilde \psi}}\right)\bar{\tilde b}_R\gamma^\mu \tilde b_R +\frac{ s_w^2}{3c_w} \bar{\tilde s}_R\gamma^\mu \tilde s_R + \left(-\frac{1}{2}+\frac{s^2_w}{3} +\frac{m^2_{\tilde b} t^2_{\theta_3}}{2m^2_{\tilde \psi}}\right)\bar{\tilde \psi}_R\gamma^\mu \tilde \psi_R  \right] \nonumber \\ 
&  +  g Z_\mu \left[ - \frac{m_{\tilde{b}}}{2 m_{\tilde \psi}} t_{\theta_3}\bar{\tilde b}_R\gamma^\mu \tilde \psi_R  
+  \frac{m_{\tilde{s}}}{2m_{\tilde \psi}} t_{\theta_2} \, \bar{\tilde s}_R\gamma^\mu \tilde \psi_R 
- \frac{m_{\tilde b} m_{\tilde s}}{2m^2_{\tilde \psi}} t_{\theta_2} t_{\theta_3} \bar{\tilde b}_R\gamma^\mu \tilde s_R \right]  + H.c. \nonumber \\
& + (b \to t, s \to c , \frac{1}{3} \to -\frac{2}{3}, -\frac{1}{2}\to \frac{1}{2} )  \,,
\label{eq:ZFermions}
\end{align} 
where the $Z$ coupling to left-handed quarks are exactly diagonal because the coupling of the new left-handed
quarks are the same as the ones of the SM left-handed quarks. 
In the second line of Eq.~(\ref{eq:ZFermions}), we see the $Z$ coupling to right-handed SM quarks
are approximately SM-like, apart from small corrections proportional to the quark masses. The third line contains 
the flavor off-diagonal interactions with SM $Z$ and right-handed quarks. 
Although suppressed by small mixing angle $t_{\theta_3}$ or by quark-mass factors, 
it is still necessary to check flavor violating processes in the top-quark sector,  e.g. $\tilde{t} \to Z \tilde{c}, Z \tilde{u} $, 
which are covered in Section \ref{sec:flavorconstraint}.

\section{Light Higgs and the CMS excess}
\label{sec:LightHiggs}

In this section, we first discuss the light Higgs $\tilde{\phi}$ with mass $96$ GeV and how it can
resolve the excess in CMS data. Later, we use the $Z^\prime$ in the same model
to fit the $R_K^{(*)}$ excess.

The dominantly production channels at LHC are gluon-gluon fusion (ggF), associate production with 
$\bar{t}t$, vector boson fusion (VBF) and associate production (VH). In the section \ref{sec:SVquarks}, 
the coupling between $\tilde{\phi}$ and quarks are given. For $m_{\tilde{\phi}} = 96$ GeV, the
coupling strength $\kappa$ is calculated, which is the coupling for $\tilde{\phi}$ devided by the coupling of
SM-like Higgs with the same mass. 
\begin{align}
& \kappa_{eff}^{\tilde t\bar{\tilde t}(\tilde b\bar{\tilde  b})\tilde \phi} =  \frac{v}{v_D} t^2_{\theta_3} \left(\frac{m^2_{\tilde t/ \tilde b}
-m^2_{\tilde \psi}c^2_{\theta_2}}{m^2_{\tilde \psi}}\right)  c_\alpha    -  \left( 1-\frac{t^2_{\theta_3}}{2}\right)s_\alpha  , \\ 
& \kappa_{eff}^{\tilde s\bar{\tilde s}(\tilde c\bar{\tilde  c})\tilde \phi} = \frac{v}{v_D}s^2_{\theta_2} c_\alpha +s_\alpha   , \\ 
& \kappa_{eff}^{VV\tilde \phi} = -s_\alpha  , \\ 
& \kappa_{eff}^{gg\tilde \phi} =  \kappa_{eff}^{\tilde t\bar{\tilde t}\tilde \phi}  + 2 \frac{v}{v_D} c_\alpha\left(s^2_{\theta_2} +\frac{t^2_{\theta_3}}{2}\left(1+c^2_{\theta_2}\right)\right)  , \\ 
& \kappa_{eff}^{\gamma\gamma\tilde \phi} =  -1.31 \kappa_{eff}^{VV\tilde \phi}  +0.31\kappa_{eff}^{\tilde t\bar{\tilde t}\tilde \phi} +0.31 \times \frac{5}{4}  \frac{v}{v_D} c_\alpha\left(s^2_{\theta_2} +\frac{t^2_{\theta_3}}{2}\left(1+c^2_{\theta_2}\right)\right)   ,
\end{align}
where $V = W,Z$. For $\kappa_{eff}^{gg\tilde \phi} $, the first term is the contribution from SM top quark,
and the second term contains the contributions from heavy vector-like quarks $\tilde{\psi}^{u,d}$
, therefore there is a factor of 2. 
For $\kappa_{eff}^{\gamma\gamma\tilde \phi} $, the third term is again contribution from 
heavy vector-like quark, with $5/4$ from $\tilde{\psi}^{u,d}$ electricmagnetic charge square
comparing to top quark. 

The ggF, and the inclusive VBF and VH production cross-sections at 13 TeV LHC are given 
below \cite{deFlorian:2016spz}
\begin{align}
& \sigma_{\text{ggF}}^{\tilde{\phi}} = 76.3 ~\text{pb} \times \left(\kappa_{eff}^{gg\tilde \phi} \right)^2   ,\\ 
& \sigma_{\text{VBF/VH}}^{\tilde{\phi}}  = 10.4 ~\text{pb}  \times s_\alpha^2 \,.
\end{align}

The dominant decay channels of $\tilde{\phi}$ have the following decay widths \cite{Heinemeyer:2013tqa}
\begin{align}
& \Gamma(\tilde \phi \to\bar {\tilde b} \tilde  b) = 1.9 ~\text{MeV} \times \left(\kappa_{eff}^{\tilde b\bar{\tilde  b}\tilde \phi}\right)^2, \quad
 \Gamma(\tilde \phi \to\tau\tau) = 0.2 ~\text{MeV} \times s^2_\alpha, \quad  
 \Gamma(\tilde \phi \to\tilde c\bar{\tilde c}) = 0.09 ~\text{MeV} \times  \left(\kappa_{eff}^{c\bar c\tilde \phi}\right)^2 \nonumber \\
& \Gamma(\tilde \phi \to gg) = 0.16 ~\text{MeV} \times\left(\kappa_{eff}^{gg\tilde \phi}\right)^2   , \quad
 \Gamma(\tilde \phi \to\gamma\gamma)  = 3.39  ~{\rm keV} \times \left(\kappa_{eff}^{\gamma\gamma\tilde \phi}\right)^2 
\end{align}

The branching ratio for $\tilde \phi $ to $\tilde b\bar {\tilde b}$ and $\gamma\gamma$ are approximately
\begin{align}
BR(\tilde \phi \to \tilde b\bar {\tilde b})& =\frac{1.9\times \left(\kappa_{eff}^{\tilde b\bar {\tilde b}\tilde \phi}\right)^2}{1.9 \times \left(\kappa_{eff}^{\tilde b\bar{\tilde  b}\tilde \phi}\right)^2 
+ 0.2 \times \sin^2\alpha + 0.16 \times\left(\kappa_{eff}^{gg\tilde \phi}\right)^2  
+ 0.09  \times  \left(\kappa_{eff}^{\tilde c\bar{\tilde  c}\tilde \phi}\right)^2} \nonumber \\
BR(\tilde \phi \to \gamma\gamma)& =\frac{3.39 \times 10^{-3} \times \left(\kappa_{eff}^{\gamma\gamma\tilde \phi}\right)^2 }{1.9 \times \left(\kappa_{eff}^{\tilde b\bar{\tilde  b}\tilde \phi}\right)^2 + 0.2\times \sin^2\alpha + 0.16  \times\left(\kappa_{eff}^{gg\tilde \phi}\right)^2  
	+0.09  \times  \left(\kappa_{eff}^{\tilde c\bar {\tilde c}\tilde \phi}\right)^2 }
\end{align}

To fit the 13 TeV CMS di-photon excess \cite{CMS:2017yta} , one needs
\begin{align}
&\sigma_{\text{ggF}}\times {\rm BR}\left(\tilde{\phi} \to \gamma\gamma \right) \sim 0.085~ {\rm  pb}.
\label{eq:CMSdi-photon}
\end{align} 
We show the parameter space $\left\lbrace \sin\alpha, \sin\theta_2 \right\rbrace$ which
fits the CMS excess in Fig.~\ref{fig:CMS-fit}, with $\sin\theta_2 \approx \lambda_2 v_D/m_{\tilde{\psi}}$. 
The  cyan solid line provides $80\%$ of CMS excess in Eq.~(\ref{eq:CMSdi-photon}), while the two dashed
line are for $60\%$ and $100\%$ of the excess respectively.
The benchmark point is denoted by a red star in the plot, and its parameters are also listed in 
Table~\ref{tab:benchmark}.  

\begin{table}[htb]
	\centering
	\begin{tabular}{|c||c|c|c|c|c|c|c|c|}
		\hline 
		model parameters&  $m_{\tilde{\phi}}$  &$\sin\alpha$  &$\sin\theta_2$& $\tan\theta_3$ & $m_{\tilde{\psi}}$ & 
		$g_D$ & $v_D$& $m_{Z^\prime}$ \\  
		\hline
		benchmark point & 96 GeV 	              & 0.1             & 0.89  &   0.1  & 800 GeV & 
		1 & 492 GeV & 4.1 TeV \\                                          
		\hline
	\end{tabular}
	\caption{The benchmark point for the signal model which fits the CMS excess and solves
		the B-anomalies simultaneously.}
	\label{tab:benchmark}
\end{table}

From Fig.~\ref{fig:CMS-fit}, the benchmark has $\sin\theta_2 = 0.89$, $m_{\tilde \psi} =\lambda_2v_D/\sin\theta_2 =800$ GeV and 
$v_D = 496$ GeV. This suggests $\lambda_2 \approx 1.4$ and the heavy quark mass dominantly comes from off-diagonal term 
$\lambda_2 v_D$. It is worth mentioning that since $\cos\theta_2 = 0.48$, the diagonal mass $m_{q^2}$
is larger than eigenstate mass $m_{\tilde{q}^2}$ due to the relationship $m_{\tilde {q}^2} = m_{q^2} \cos\theta_2$. 

For the benchmark point, we list the 96 GeV Higgs branching ratios in Table \ref{tab:phiBR}. 
\begin{table}[!h]
	\begin{center}
		\begin{tabular}{|c||c|c|c|c|c|c|c|c|c|c|}
			\hline 
			process&  $b\bar{b}$  &$\tau\bar{\tau}$ & $c\bar{c}$ & $gg$ & $\gamma\gamma$ & $WW^*$ &$ZZ^*$ & total \\  \hline
			\hline
			BR & $15.9\%$ & $1.66\%$  & $18.23\%$  & $63.9\%$&   $1.8\times10^{-3}$ & $8.3\times10^{-4}$  & $1.41\times10^{-4}$ & 1\\       
			\hline
			$ \Gamma_i$ (MeV)& 0.019 & 0.022  & 0.002  & 0.077&   $2.18\times10^{-4}$ & $1\times 10^{-4}$  & $1.7\times10^{-5}$& 0.12 \\                                        
			\hline
		\end{tabular}
	\end{center}
	\caption{The decay widths and branching ratios  for a 96 GeV $\tilde{\phi}$ and parameters set at the benchmark
	point values, given in Table~\ref{tab:benchmark}.	
	}
	\label{tab:phiBR}
\end{table}

\subsection{Constraints from SM Higgs measurements}
\label{sec:Hconstraint}

In this subsection, we are going to check the limits from SM Higgs measurements. For convenience,
we list the coupling strength $\kappa$ for the SM-like Higgs $\tilde{h}$ below,
\begin{align}
&\kappa_{eff}^{\tilde t\bar{\tilde t}(\tilde b\bar{\tilde  b})\tilde h} =  \frac{v}{v_D} t^2_{\theta_3} \left(\frac{m^2_{\tilde t/ \tilde b}-m^2_{\tilde \psi}c^2_{\theta_2}}{m^2_{\tilde \psi}}\right)  s_\alpha    + \left( 1-\frac{t^2_{\theta_3}}{2}\right)c_\alpha,  \\ 
& \kappa_{eff}^{\tilde s\bar{\tilde s}(\tilde c\bar{\tilde  c})\tilde h} = \frac{v}{v_D}\tan^2\theta_2 \sin\alpha -\cos\alpha, \\ 
& \kappa_{eff}^{VV\tilde h} = \cos\alpha, \\ 
& \kappa_{eff}^{gg\tilde h} =  \kappa_{eff}^{\tilde t\bar{\tilde t}\tilde h}  + 2 \frac{v}{v_D} s_\alpha\left(s^2_{\theta_2} +\frac{\tan^2\theta_3}{2}\left(1+c^2_{\theta_2}\right)\right),  \\ 
& \kappa_{eff}^{\gamma\gamma\tilde h} =  -1.31 \kappa_{eff}^{VV\tilde h}  +0.31\kappa_{eff}^{\tilde t\bar{\tilde t}\tilde h} +0.31 \times \frac{5}{4}  \frac{v}{v_D} s_\alpha\left(s^2_{\theta_2} +\frac{\tan^2\theta_3}{2}\left(1+c^2_{\theta_2}\right)\right) .   
\end{align}

\begin{figure}[htb]
	\centering
	\includegraphics[width=0.48 \textwidth]{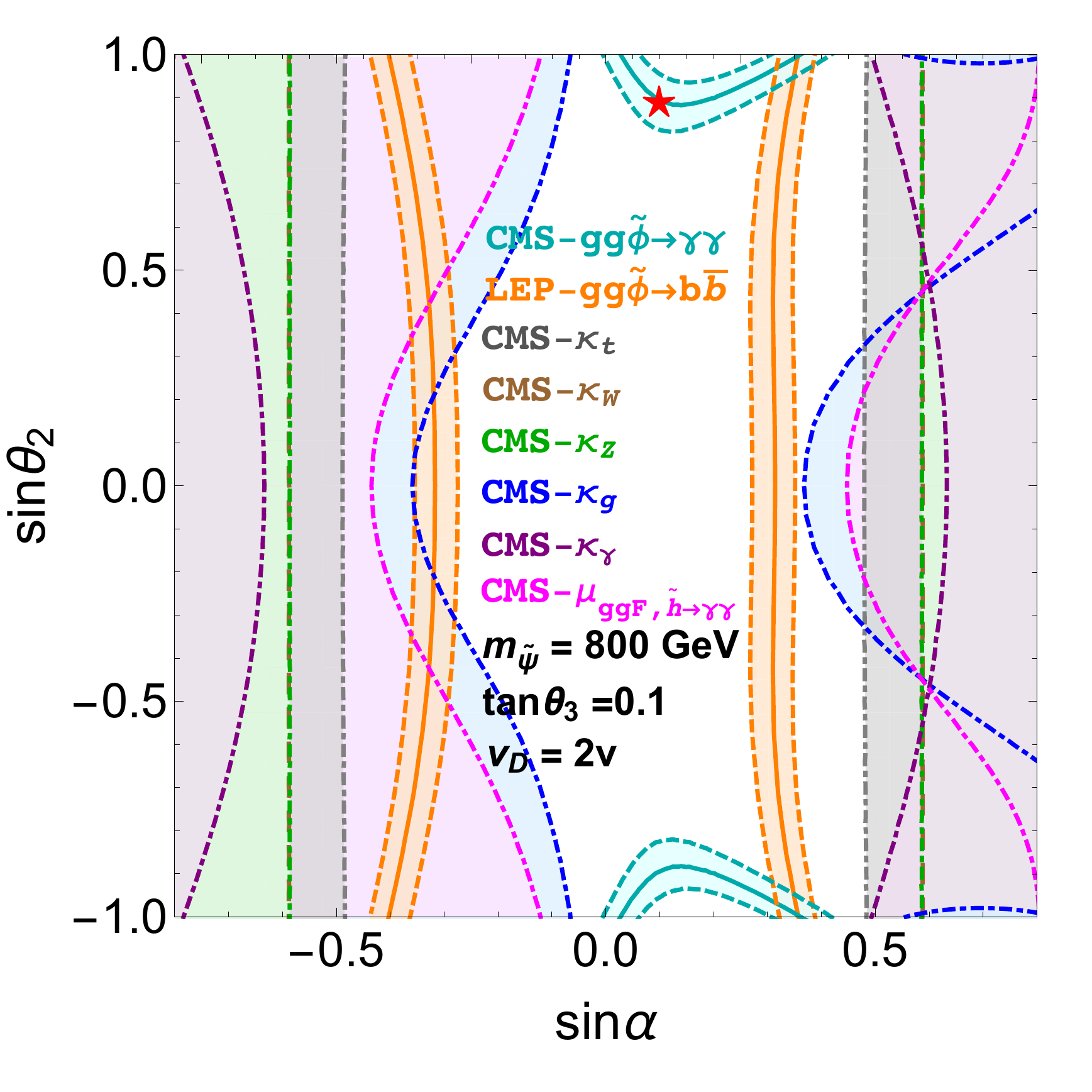} 
	\includegraphics[width=0.48 \textwidth]{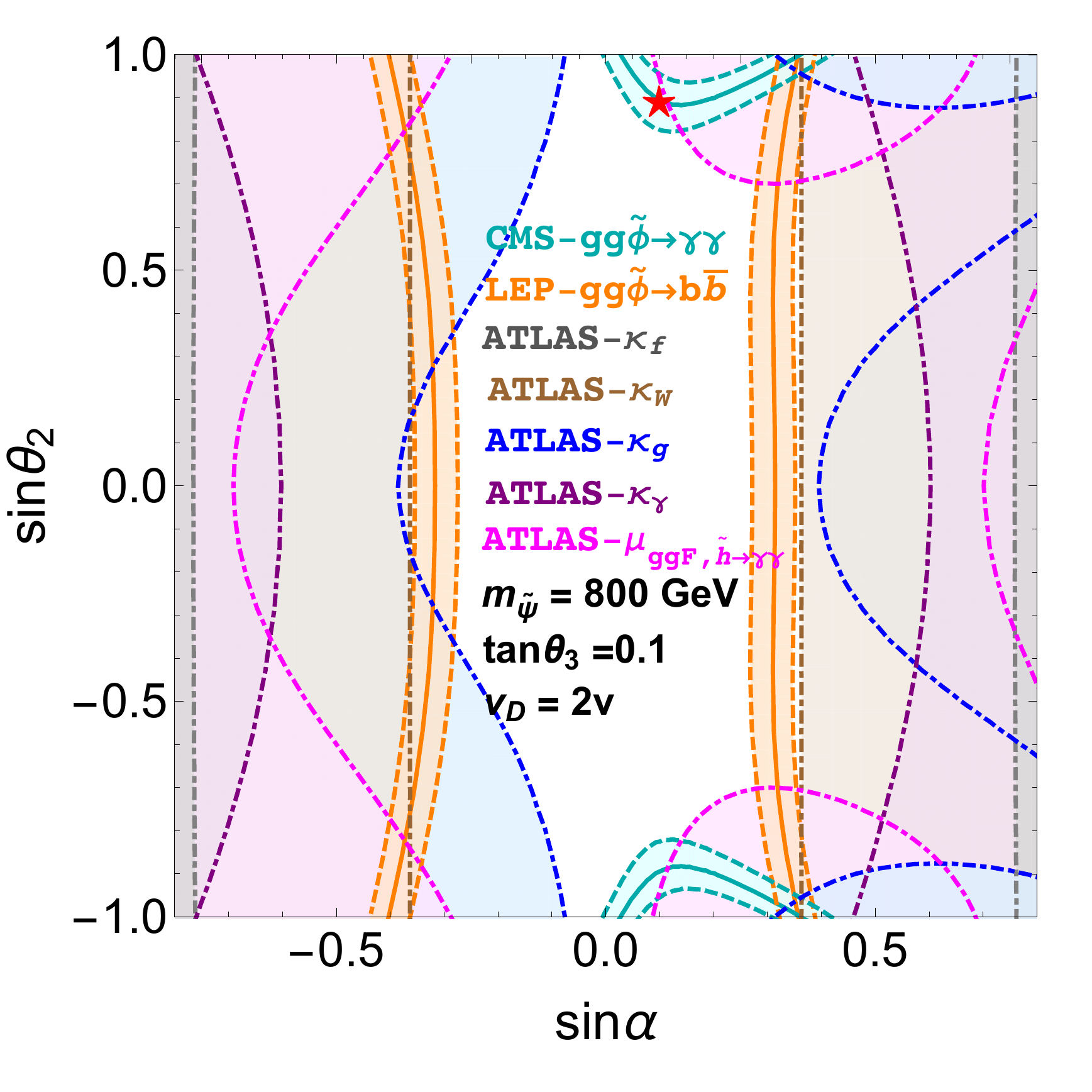} 
	\caption{ The $90\%$ C.L. constraints from  SM-like Higgs precision measurements and the parameter space fitting
		the CMS \cite{CMS:2017yta} (``$gg\tilde{\phi}\to \gamma\gamma$") and LEP (``$e^+e^- \to Z\tilde{\phi}\to Z \bar{b}b$")\cite{Barate:2003sz} signals are plotted in $\sin\alpha-\sin\theta_2$ 2D plane with cyan and orange color shading denoting the signal regions. All the other color shaded regions
		are excluded from various experiments. 
		All the constraints from CMS (ATLAS) are in the left (right) panel respectively. The constraints from SM-like Higgs coupling 
		strength, denoted as ``$\kappa$", are from CMS~\cite{CMS-PAS-HIG-17-031} and ATLAS~\cite{ATLAS-CONF-2017-047} 
		measurements at 13 TeV ($36 ~\text{fb}^{-1}$). The SM Higgs di-photon decay analysis in the gluon fusion production
		give signal strength $\mu_{\text{ggF}, h \to \gamma\gamma}$ limits for CMS \cite{Sirunyan:2018ouh} and ATLAS \cite{Aaboud:2018xdt}.
		The $90\%$ C.L. exclusion regions are colored with  dot-dashed boundaries. 
		The parameter region consistent with the  CMS (``$gg\tilde{\phi}\to \gamma\gamma$") and LEP (``$e^+e^- \to Z\tilde{\phi}\to Z\bar{b}b$") signals 
		are   the same in the left and right panels, and denoted by  cyan and orange colors, respectively. The solid line shows parameters
		leading to an explanation of  $80\%$ of these signals, while 
		the two dashed lines are consistent with an explanation of $60\%$ and $100\%$ of these signals, respectively. The red star denotes the parameters associated with 
		the benchmark point shown in Table \ref{tab:benchmark}.
	 }
	\label{fig:CMS-fit}
\end{figure}

The most recent 13 TeV ($36 ~\text{fb}^{-1}$) constraints on SM Higgs coupling strength $\kappa$ are from 
CMS combined measurement \cite{CMS-PAS-HIG-17-031}, 
\begin{align}
&\kappa_{t}^{\rm CMS13}=1.09^{+0.14}_{-0.14} , \quad
\kappa_{W}^{\rm CMS13}=1.12^{+0.13}_{-0.19} , \quad
\kappa_{Z}^{\rm CMS13}=0.99^{+0.11}_{-0.11} , \nonumber \\
&\kappa_{g}^{\rm CMS13}=1.14^{+0.15}_{-0.13} , \quad
\kappa_{\gamma}^{\rm CMS13}=1.07^{+0.15}_{-0.18} .
\label{eq:CMSkappa}
\end{align}
This CMS analysis includes ggF, VBF, VH and ttH productions, and various SM Higgs decay modes. 
For ATLAS, the production modes are the same, but only $H\to \gamma\gamma$ and $H \to Z Z \to 4\ell$ decay modes 
are included in the $\kappa$ analysis \cite{ATLAS-CONF-2017-047},
\begin{align}
&\kappa_{f}^{\rm ATLAS13}= 0.89^{+0.20}_{-0.15} , \quad 
\kappa_{V}^{\rm ATLAS13}=1.03 \pm 0.06 , 
\nonumber \\
& \kappa_{g}^{\rm ATLAS13}=1.08^{+0.11}_{-0.10} , \quad
\kappa_{\gamma}^{\rm ATLAS13}=0.93^{+0.09}_{-0.08} .
\label{eq:ATLASkappa}
\end{align}  
It is clear that $\kappa$ from CMS and ATLAS are in agreement with each other.
We took the $\kappa$ measurements as constraints and plot the corresponding contours at $90\%$ confidence level 
(C.L.) in Fig.~\ref{fig:CMS-fit}, with CMS in the left panel and ATLAS in the right panel.

Since in our signal model $\tilde{\phi}$ is produced from ggF and decay in di-photon channel, it is appropriate 
to pay special attention in the SM Higgs di-photon channel measurement as well. The most recent
13 TeV ($36 ~\text{fb}^{-1}$) measurements for SM Higgs property in the di-photon decay channel, 
from CMS \cite{Sirunyan:2018ouh} and ATLAS \cite{Aaboud:2018xdt} give the following signal strengths for different 
production modes and the combined result,
\begin{align}
\mu_{ggF, h \to \gamma\gamma}^{\rm CMS13}=1.10^{+0.20}_{-0.18} 
, \quad \mu_{VBF, h \to \gamma\gamma}^{\rm CMS13}=0.8^{+0.6}_{-0.5}
, \quad \mu_{ttH, h \to \gamma\gamma}^{\rm CMS13}=2.2^{+0.9}_{-0.8}   
,  \quad \mu_{\text{combined}, h \to \gamma\gamma}^{\rm CMS13}=1.18^{+0.17}_{-0.14}   
\\
\mu_{ggF, h \to \gamma\gamma}^{\rm ATLAS13}=0.81^{+0.19}_{-0.18} 
, \quad \mu_{VBF, h \to \gamma\gamma}^{\rm ATLAS13}=2.0^{+0.6}_{-0.5}
, \quad \mu_{ttH, h \to \gamma\gamma}^{\rm ATLAS13}=0.5^{+0.6}_{-0.6}   
,  \quad \mu_{\text{combined}, h \to \gamma\gamma}^{\rm ATLAS13}=0.99^{+0.15}_{-0.14}   .
\label{eq:signalstrengths}
\end{align}
The above measurements in the di-photon channel have improved the precision by factor of 2 compared to the ones  with
Run-I data \cite{Aad:2014eha}, therefore we do not consider the Run-I constraints further. 
The most relevant and stringent constraint from the signal strengths is from $\mu_{ggF, h \to \gamma\gamma}$. 
The ATLAS result is barely consistent with SM at $1~ \sigma$. Thus, we give $90\%$ C.L. ($1.64 ~\sigma$) constraints for 
CMS and ATLAS respectively in Fig.~\ref{fig:CMS-fit}. The excluded parameter regions are plotted as magenta color with
dot-dashed boundary.

\subsection{Simultaneous explanation of the CMS and LEP excesses.}

In Fig.~\ref{fig:CMS-fit}, we show the constraints from Higgs precision measurements, as well as the region consistent with the LEP excess,
shown as the orange regions in both left and right panels of Fig.~\ref{fig:CMS-fit}.  While the CMS excess can be easily explained without 
being in tension with any current Higgs measurements, the simultaneous explanation of the CMS di-photon and LEP $b\bar{b}$ excesses can be
only obtained for values of $s_{\theta_2} \simeq \pm 1$ and $s_\alpha \simeq  0.3$.  It is easy to see that, for these values of the mixing 
angles $\kappa^{gg\tilde{h}}_{eff} \simeq 1.2$.  

The large $\kappa^{gg\tilde{h}}_{eff}$ induces a sizable rate for the SM-like Higgs production in the gluon fusion channel.  While
such an enhancement is consistent with the current CMS measurement in the ggF channel, both in the di-photon as well as in the $ZZ$
and $WW$ final states, it is in tension with the current ATLAS measurement of the ggF production of SM-like Higgs bosons in 
the di-photon channel, see Eq.~(\ref{eq:signalstrengths}). This tension is clearly shown in the right  panel of Fig.~\ref{fig:CMS-fit}.

In summary, while the LEP and CMS excesses can be easily separately explained within this framework, a simultaneous explanation of the CMS and 
LEP excesses is in tension with the current ATLAS observations in the ggF channel. In the following, we shall concentrate on the benchmark model,
which leads to an explanation of the CMS di-photon excess, and ignore the LEP $\bar{b} b$ excess.  However, beyond the SM-like 
Higgs physics, all other phenomenological aspects of this model would be only minimally affected by a change from the benchmark 
point parameters to the region of parameters, where the LEP and CMS excesses are simultaneously explained.

\section{$Z'$ and the B-anomalies}
\label{sec:BAnomalies}

\begin{figure}
	\centering
	\includegraphics[width=0.48 \textwidth]{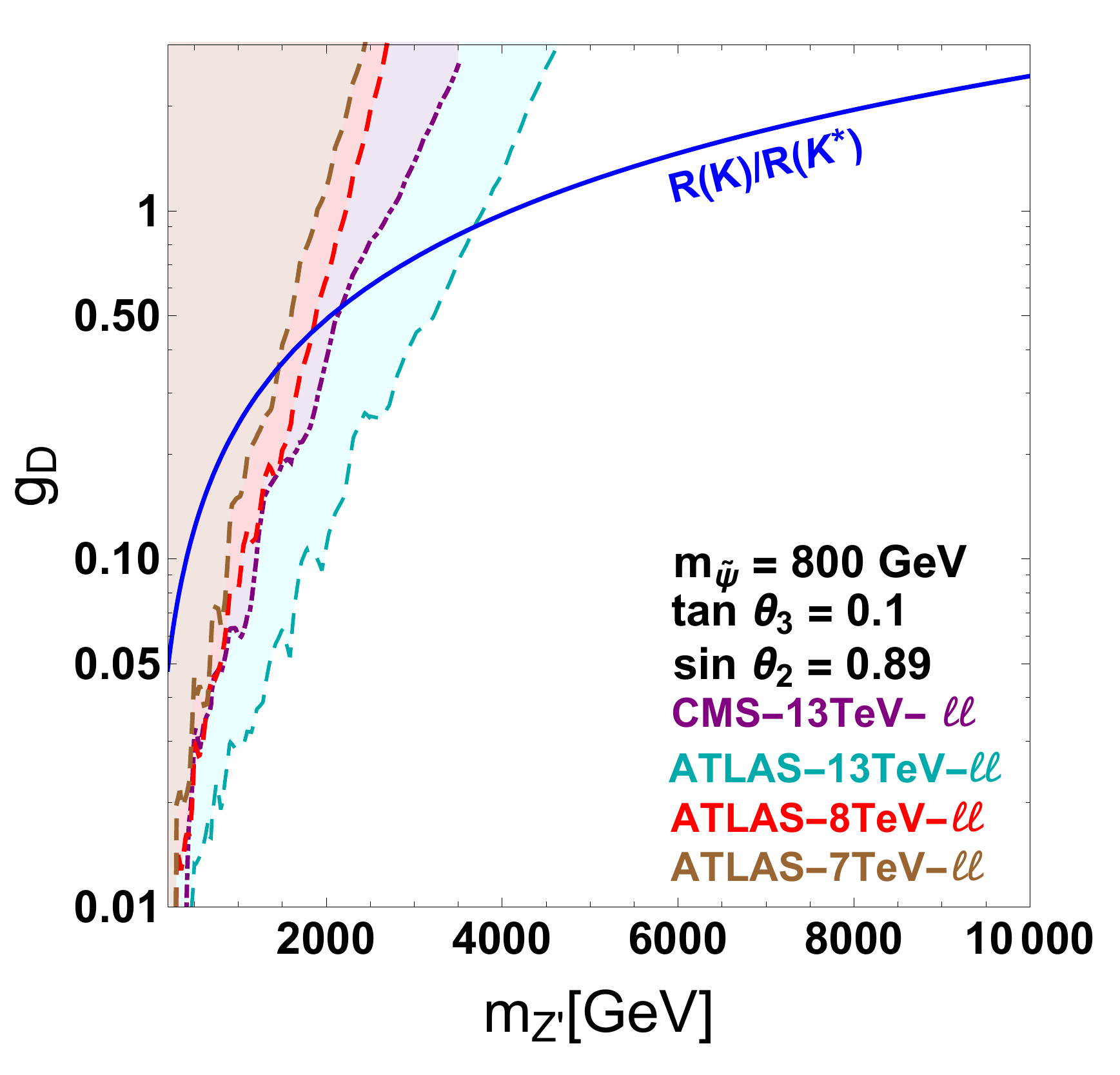} 
	\includegraphics[width=0.48 \textwidth]{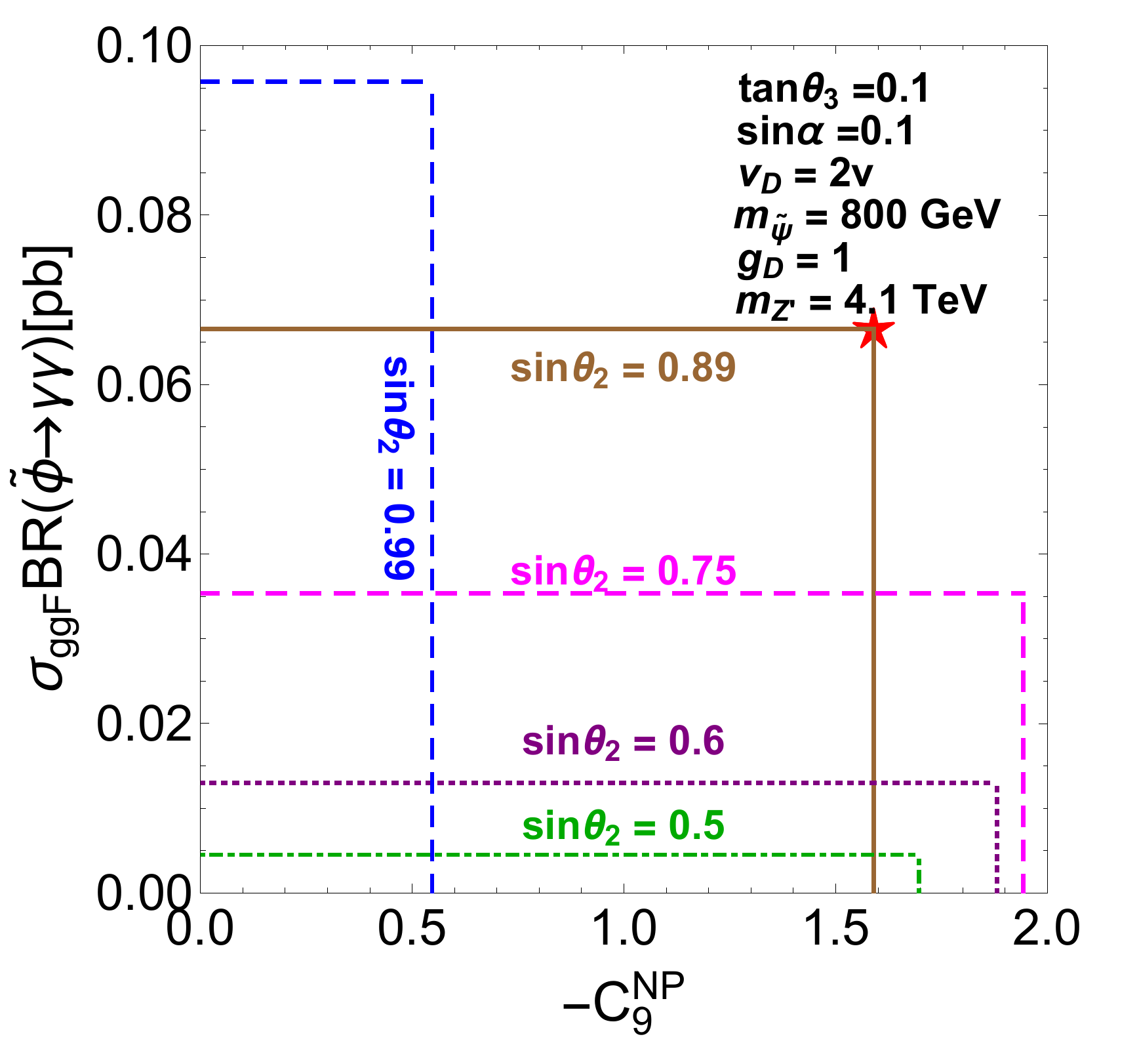} 
	\caption{ Left panel: the constraints on $Z'$ in $m_{Z'}-g_D$ 2D plane. The colored regions are excluded by the LHC search
		for heavy neutral gauge boson. The purple area is excluded by CMS 13 TeV search on di-leptons \cite{Khachatryan:2016zqb}. 
		The cyan and red areas are excluded by ATLAS 13 TeV \cite{Aaboud:2016cth,Aaboud:2017buh} and 8 TeV di-lepton 
		search \cite{Aad:2014cka}. The brown area is excluded by 7 TeV ATLAS search on di-leptons \cite{Aad:2012hf}. 
		The blue line satisfies the requirement from B-anomalies. Right panel: Contours of constant values of  $\sin \theta_2$,
		showing the dependence on this mixing angle of  the  coefficient
		  $-C_9^{NP}$  and the di-photon signal cross-section $\sigma_{\text{ggF}}  \text{BR}(\tilde{\phi} \to \gamma\gamma)$ at
		   the	13 TeV LHC. All the parameters except  from $\sin\theta_2$ are taken from Table \ref{tab:benchmark}, and the red star 
		   denotes our benchmark point. } 
	\label{fig:Zp-constraint}
\end{figure}

After discussing the possibility of $\tilde{\phi}$ fitting the CMS excess, we turn to the B-anomalies. 
Integrating out the heavy gauge boson $Z^\prime$, there is an effective flavor violating operator with
down-type quarks \cite{Altmannshofer:2014cfa},
\begin{align}
\mL_{eff} = -\frac{g_D^2}{2 m^2_{Z'}} s_{\theta_2} c_{\theta_2} t_{\theta_3}\bar {\tilde b}_L \gamma_\mu \tilde s_L 
\bar {\mu} \gamma^\mu \mu  + H.c. \,.
\end{align}
This can be related to the $C_9^{NP}$ operator considered in Ref.~\cite{Altmannshofer:2013foa,Altmannshofer:2014rta,
Altmannshofer:2015sma,Greljo:2015mma, Descotes-Genon:2015uva,Hurth:2016fbr,Altmannshofer:2017yso,
Capdevila:2017bsm,Altmannshofer:2017fio}:
\begin{align}
\label{eq:Heff}
\mH_{eff}^{NP} = -\mL_{eff}^{NP}  = - \frac{4 G_F}{\sqrt{2}}\frac{\alpha_{em}}{4\pi}( V_{tb} V^*_{ts}) C_9^{NP}   
\bar {\tilde b}_L \gamma_\mu \tilde s_L \, \bar {\mu} \gamma^\mu \mu  + H.c. \,,
\end{align}
therefore, the coefficient $C_9^{NP}$ can be rewritten as
\begin{align}
C_9^{NP} =  -\frac{ g_D^2 v^2}{ m_{Z^\prime}^2} \frac{\pi}{\alpha_{em}} \frac{1}{V_{tb}V_{ts}^*} 
s_{\theta_2}c_{\theta_2}t_{\theta_3}  \,.
\label{eq:C9NP}
\end{align}
Recent global fits include more data from experiments, e.g. angular observables from Belle \cite{Wehle:2016yoi}, and have 
found that the significance of NP contributions has increased~\cite{Capdevila:2017bsm, Altmannshofer:2017yso}. 
If one restricts the analysis to lepton flavor universality violation process, 
the value $C_9^{NP} =-1.56$ quoted above was obtained by a fit to the data by the authors of Ref.~\cite{Altmannshofer:2017yso}, 
with a significance of 4.1~$\sigma$,
while the authors of Ref.~\cite{Capdevila:2017bsm} obtained a slightly different result, namely, $C_9^{NP} =-1.76$,
with a significance of 3.9~$\sigma$. However, extending the analysis to a more complete set of observables, namely all
those included in Ref.~\cite{Capdevila:2017bsm}, a best fit value of $C_9^{NP} = -1.11$ is obtained, and the significance increases to 5.8~$\sigma$.
In our analysis, we shall consider the values obtained from the fit in Ref.~\cite{Altmannshofer:2017yso}.  The alternative
values of $C_9^{NP}$ obtained in Ref.~\cite{Capdevila:2017bsm}  do not affect our phenomenological analysis in any significant way, 
since they  can be easily accommodated by few tens of percent changes in the mass of the gauge boson $Z'$. 
After plugging in the SM parameters ($\alpha_{em} = 1/137, |V_{ts}| \sim 0.04$), we obtain a requirement on NP parameters,
\begin{align}
\frac{g_D^2}{m_{Z'}^2} s_{\theta_2}c_{\theta_2}t_{\theta_3}  \sim 2.44 \times 10^{-3} {\rm TeV}^{-2} \,.
\label{eq:RKrequirement}
\end{align}
For the explanation of the $R_K$ and $R^*_K$ anomalies, we have chosen a benchmark with $t_{\theta_3} = 0.1$ 
and $s_{\theta_2} = 0.89$, which are consistent with the parameters shown by the red star in  Fig.~\ref{fig:CMS-fit}. 
For such a benchmark, one obtains a relationship between $g_D$ and $m_{Z'}$ given by
$\frac{m_{Z'}}{g_D} \simeq 4.1 ~{\rm TeV}$, which is shown as the blue line in the left panel of Fig. \ref{fig:Zp-constraint}.
In the right panel of Fig. \ref{fig:Zp-constraint}, we show contours of constant values of  $\sin \theta_2$,
		showing the dependence on this mixing angle of  the  coefficient
		  $-C_9^{NP}$  and the di-photon signal cross-section $\sigma_{\text{ggF}}  \text{BR}(\tilde{\phi} \to \gamma\gamma)$ at
		   the	13 TeV LHC.  We see that di-photon signal at CMS requires a large value of $\sin \theta_2$, while the explanation
of the B-anomalies requires a moderate value of $\sin \theta_2$ since $-C_9^{NP}$ is proportional to $s_{\theta_2} c_{\theta_2}$.

\subsection{Constraints on $Z'$}
\label{sec:Zpconstraint}

In Eq.~(\ref{eq:ZpFermions}), we have already presented the interactions of $Z'$ and in Fig. ~\ref{fig:Zp-BR}
we show the branching ratios of $Z'$. We find out that, depending on the $Z'$ mass, the dominant decay channel of $Z'$ 
are $Z' \to \bar{\tilde s} \tilde{s}  ~( \bar{\tilde c} \tilde{c})$, $Z'\to \bar{\tilde \psi}^u \tilde{\psi}^u ~(\bar{\tilde \psi}^d \tilde{\psi}^d)$, 
$Z'\to \bar{\mu}  \mu ~ (\bar{\tau}\tau)$ and $Z' \to \nu_{\mu} \bar\nu_\mu ~(\nu_{\tau} \bar\nu_\tau)$. 

We now consider the constraints from $Z'$ searches at LHC. The search for new neutral gauge bosons decaying to di-lepton~\cite{Aad:2012hf,Aad:2014cka,Aaboud:2016cth,Khachatryan:2016zqb,Aaboud:2017buh} and di-jet \cite{CMS:2017xrr} 
put strong constraints on a possible $Z'$. After consider the $Z'$ production and decay branching ratios, in 
in the left panel of Fig.~\ref{fig:Zp-constraint},
we show the constraints in $g_D$ vs. $m_{Z'}$ 2D plane with benchmark parameters $\tan\theta_3=0.1$, $\sin\theta_2=0.89$ and 
$m_{\tilde \psi} = 800$ GeV.  All the colored regions are excluded. The limits from di-jet are much weaker than di-lepton, because
$Z'$ couples to SM quark only through heavy vector-like quark mixing while leptons are directly charged. Thus di-jet constraint 
is not shown in Fig.~\ref{fig:Zp-constraint}. The solution to B-anomalies requires $m_{Z'}/g_D \simeq 4.1$ TeV. It is plotted 
as a solid blue line and denoted as ``$R(K)/R(K^*)$" in Fig.~\ref{fig:Zp-constraint}. To evade the constraints,
$Z'$ must be heavier than about 4.1~TeV, while $g_D$ must be larger than about 0.8.

For $U(1)_{L_\mu -L_\tau}$ gauge boson, it is known that neutrino trident production $\nu_{\mu } N \to \nu_{\mu } N \mu^+\mu^-$, 
where $N$ is nuclei, gives very stringent constraint \cite{Altmannshofer:2014cfa, Altmannshofer:2014pba}. The recent
results are from CHARM-II \cite{Geiregat:1990gz}, CCFR \cite{Mishra:1991bv} and NuTeV \cite{Adams:1998yf}, 
which leads to constraint $m_{Z^\prime}/g_D \gtrsim 540$ GeV for $m_{Z^\prime} \gtrsim 10$ GeV \cite{Fuyuto:2015gmk}.
It is easy to see that our benchmark point for model is safe from neutrino trident constraint. Observe that a simultaneous explanation of the
B-anomalies and the LEP bottom forward-backward asymmetry  with a single $Z'$ is not possible, since the latter demands a light
$Z'$, with mass comparable to the $Z$ mass and significant couplings to the right-handed bottom quarks~\cite{Liu:2017xmc}, what 
would lead to  inconsistencies with the above constraints.

\section{Other constraints}
\label{sec:Constraints}

\subsection{Constraints on the heavy quarks}
\label{sec:psiconstraint}
\begin{figure}
	\centering
	\includegraphics[width=0.5 \textwidth]{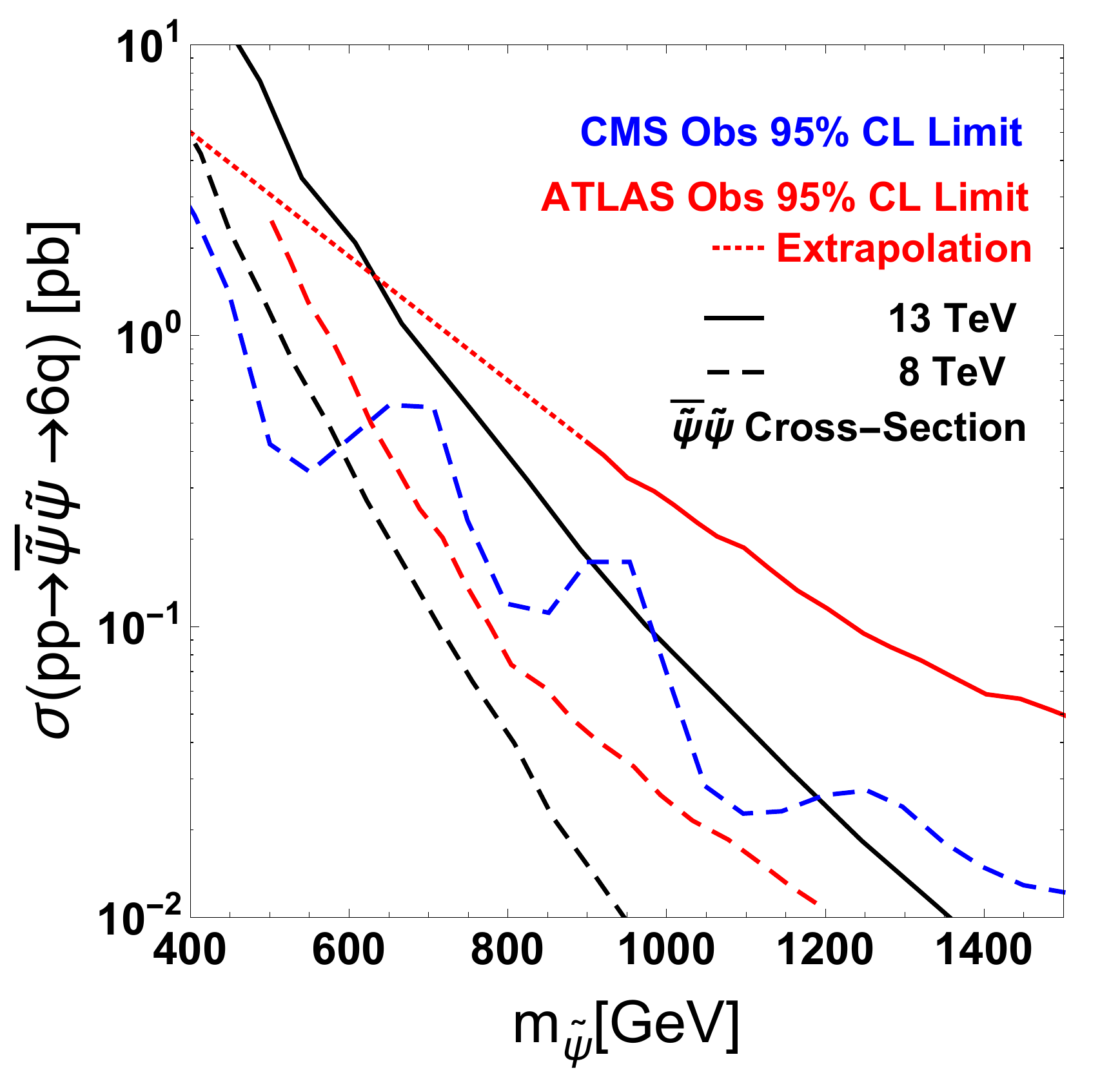} 
	\caption{ The $95\%$ C.L. constraints on the heavy vector-like quark pair production cross-section with subsequent 
		decay into three light flavor jets. The $95\%$ C.L. limits are from RPV gluino search at 13 TeV ATLAS \cite{Aaboud:2018lpl} 
		and at 8 TeV CMS \cite{Chatrchyan:2013gia} and ATLAS \cite{Aad:2015lea}. The 13 (8) TeV limits are plotted as solid 
		(dashed) lines, and ATLAS and CMS are using red and blue color respectively. The black lines show the pair
	production cross-sections for $\tilde{\psi}$ at 8 TeV and 13 TeV.}
	\label{fig:RPV-Limit}
\end{figure}

The heavy vector-like quark is subject to constraints from LHC searches. Its mass is  $
m_{\tilde \psi} \simeq \sqrt{\lambda_2^2 v_D^2   + m_\psi^2} $, and it is about $m_{\tilde \psi} \simeq 800 $ GeV for the benchmark point. 
At the LHC, such a quark would be produced by QCD processes and will predominantly decay into the
$\tilde \psi \to \tilde{\phi} \tilde s/\tilde c$ channels, followed by $\tilde{\phi} \to \bar {\tilde c} \tilde c, \bar {\tilde s} \tilde s , \bar {\tilde b} \tilde b$.
There are other possible decay channels, e.g. $\tilde \psi^q \to \tilde{\phi} q^3$ and $\tilde{\psi}^q \to \tilde{h} q^2$ however suppressed
by either $t_{\theta_3}^2$ or $s_\alpha^2$, and $\tilde{\psi} \to Z q$ suppressed by small quark mass or $t_{\theta_3}$ 
(see Eq.~(\ref{eq:ZFermions})). The decay $\tilde \psi \to Z' \tilde{s}/\tilde{c}$ is also possible but is kinematically forbidden by
the benchmark setup $m_{Z'}> m_{\tilde \psi}$. Therefore, the dominant constraints come from searches for heavy fermions
which decay into three quarks. A recent relevant search is from ATLAS 13 TeV ($36~\text{fb}^{-1}$) looking for R-Parity violating (RPV)
gluino at the LHC, which decays to three quarks~\cite{Aaboud:2018lpl}. It sets $95\%$ C.L. upper limit on the cross section times branching ratio 
varies between $0.80$ fb at gluino mass 900 GeV and $0.011$ fb at gluino mass 1800 GeV. At LHC Run-I, ATLAS \cite{Aad:2015lea}
and CMS \cite{Chatrchyan:2013gia} have a similar limit for RPV gluino at 8 TeV ($20 ~\text{fb}^{-1}$), which excludes gluino mass lower 
than 650 GeV if it decays into three light-flavor jets. In Fig.~\ref{fig:RPV-Limit}, we show the pair production cross-section of $\bar{\tilde{\psi}} \tilde{\psi}$
at 13 TeV and 8 TeV \cite{ElHedri:2017nny,Czakon:2013tha}, including both $\tilde{\psi}^u$ and $\tilde{\psi}^d$. 
The limits on $\sigma \times BR$ from ATLAS and CMS are plotted as red and blue lines, while 13 TeV and 8 TeV are using solid and dashed
lines respectively. The 13 TeV ATLAS data has stopped at 900 GeV and we extrapolate it down to 400 GeV with dotted line.
From Fig.~\ref{fig:RPV-Limit}, it is clear that $\tilde{\psi}$ mass larger than 600 GeV are compatible with the 8 TeV constraints
and ATLAS 13 TeV data requires $\tilde{\psi}$ mass larger than 650 GeV if the extrapolation is correct. 
The above limits are valid for $\tilde{\psi}$ decaying to three light flavor jets. The CMS 8 TeV search \cite{Chatrchyan:2013gia}
has also set limit if it decays to $qqb$, and the constraint is slightly better than $qqq$ decay channel. The ATLAS 8 TeV
search \cite{Aad:2015lea} has looked for $bbb$ decay channel, and the constraint is roughly the same as $qqq$ channel.
The heavy vector-like quark $\tilde{\psi}$ can decay to $qbb$, if $\tilde{\phi} \to \bar{b}b$. Such decay does not match to
three jets invariant mass reconstruction for either $qqb$ or $bbb$, therefore more possible combinatorial errors will reduce
the signal efficiency. Moreover, it will pay double suppression from branching ratio $\tilde{\phi} \to \bar{b}b$. As a result,
$\tilde{\psi}$ decays into three light flavor jets are more general and useful. In summary, our benchmark model is not excluded
by the three jet resonance searches at LHC.

\subsection{Constraints from flavor physics}
\label{sec:flavorconstraint}

We start the discussion from $b \to s \gamma$ constraints. The SM branching ratio for $b\rightarrow s\gamma$ is given by \cite{Barbieri:1993av}:
\beq
\frac{\text{BR} (b\rightarrow s\gamma)}{\text{BR} ( b\rightarrow c e \bar{\nu}_e)} = \frac{6\alpha}{\pi}\frac{\left[\eta^{\frac{16}{23} }A_\gamma 
+ \frac83 (\eta^{\frac{14}{23}} - \eta^{\frac{16}{23}}) A_g + C \right]^2}{I(m_c/m_b)\left[1 - \frac{2}{3\pi} \alpha_s(m_b) f(m_c/m_b)\right]} \,,
\eeq
where $\eta = \frac{\alpha_s(m_Z)}{\alpha_s(m_b)} = 0.548$, $I(m_c/m_b) = 0.58$ is the phase-space factor and $C = -0.18$ is a coefficient 
from the operator mixing. The Wilson coefficients $A_\gamma (m_t^2/m_W^2), A_g(m_t^2/m_W^2)$ are defined as following:
\beq
\mL_{eff} = G_F \frac{\alpha}{8 \pi^3} V_{tb} V^*_{ts} m_b \left[ A_\gamma \bar{s}_L \sigma^{\mu\nu} b_R F_{\mu\nu} + A_g \bar{s}_L \sigma^{\mu\nu} T^a b_R G_{\mu\mu}^a\right] + H.c. \,,
\eeq
where $F_{\mu\nu}, G_{\mu\nu}$ are the photon and gluon field strengths.  The SM CKM matrix values are given by \cite{Patrignani:2016xqp}:
\beq
|V_{tb}| = 1.009 \pm 0.031, \qquad |V_{ts}| = 0.04 \pm 0.0027 \,.
\eeq 
For the formula for $A_{\gamma,g}$ from the SM $W$-loop, we refer to the appendix of Ref.~\cite{Barbieri:1993av}. In our model, the 
contributions to $b\rightarrow s \gamma$ decay come from $Z^\prime,Z$ flavor off-diagonal couplings in \Eq{eq:ZpFermions} and 
\Eq{eq:ZFermions}.  We are not going to calculate the contributions directly, but only make an estimation for the order of magnitude 
comparing with the contribution from $W$-loop. It is not difficult to estimate their contribution to the Wilson coefficients $A_\gamma, A_g$ :
\begin{align}
A_{\gamma,g}^{Z} &\sim \frac{1}{V_{tb}V_{ts}^*} \frac{m_W^2}{m_Z^2} \frac{m_{\tb} m_{\ts} }{ m_{\tpsi}^2 } \left[ A_{\gamma,g}^{\rm SM}(m_{\tb}^2/m_Z^2) \frac{s_w^2}{6 c_w  } + A_{\gamma,g}^{\rm SM}(m_{\ts}^2/m_Z^2) \frac{s_w^2}{6 c_w  }  + \frac14 A_{\gamma,g}^{\rm SM}(m_{\tpsi}^2/m_Z^2) \right]  , \\
A_{\gamma,g}^{Z^\prime} &\sim \frac{1}{V_{tb}V_{ts}^*} \frac{g_D^2 m_W^2}{4 g^2m_{Z^\prime}^2} t_{\theta_3} c_{\theta_2} s_{\theta_2}\left[ A_{\gamma,g}^{\rm SM}(m_{\tb}^2/m_{Z'}^2) t_{\theta_3}^2 c_{\theta_2}^2+ A_{\gamma,g}^{\rm SM}(m_{\ts}^2/m_{Z'}^2) s_{\theta_2}^2 +  A_{\gamma,g}^{\rm SM}(m_{\tpsi}^2/m_{Z'}^2)  c_{\theta_2}^2\left(\frac{t_{\theta_3}^2}{2} -1\right)   \right] \,. \nonumber
\end{align}
From the equation above, we can see that the contribution is either suppressed by $m_{\tilde b} m_{\tilde s}/m_{\tilde \psi}^2$ or $m_W^2/m_{Z'}^2$, 
which leads to negligible contribution for the $b\rightarrow s \gamma$ decay. To be more explicit, we find that for 
$m_{Z'} \sim 4.1  \TeV, g_D \sim 1, m_{\tpsi} \sim 800 \GeV$ and $t_{\theta_3} \ll 1$, we have:
\beq
\frac{A_{\gamma}^{Z} }{A_\gamma^{SM}} \sim 5 \times 10^{-6} , \quad \frac{A_{g}^{Z} }{A_g^{SM}} \sim 4 \times 10^{-6} , \quad  \frac{A_{\gamma}^{Z'} }{A_\gamma^{SM}} \sim 3 \times 10^{-4}  \, t_{\theta_3} c_{\theta_2}^3 s_{\theta_2}, \quad \frac{A_{g}^{Z'} }{A_g^{SM}} \sim  4 \times 10^{-4}  \, t_{\theta_3} c_{\theta_2}^3 s_{\theta_2},
\eeq
which are very small and do not induce any relevant modification to the $b\rightarrow s \gamma$ decay rate.  Note that the operator which explains the lepton-flavor university violation in \Eq{eq:Heff} will not contribute to $B_s \rightarrow \mu^+ \mu^-$ decay, as our $Z^\prime$ couplings to muon pair is vector-like~\cite{Buras:2012jb}.

Apart from the operator $\bar{\tilde{b}}_L \gamma^\mu \tilde{s}_L\bar{\mu}\gamma_\mu \mu$, there are also $\Delta F = 2$ flavor changing 
operators generated:
\beq
\mL_{eff} = - \frac{g_D^2}{4m_{Z^\prime}^2} s_{\theta_2}^2  c_{\theta_2}^2   t_{\theta_3}^2\left(\bar{\tb}_L \gamma^\mu \ts_L\right)^2 - \frac{g^2}{4 m_Z^2} \frac{m_{\tilde b}^2 m_{\tilde s}^2 }{m_{\tilde \psi}^4} t_{\theta_2}^2 t_{\theta_3}^2\left(\bar{\tb}_R \gamma^\mu \ts_R\right)^2 + H.c. \,.
\eeq
Both operators will contribute to $\bar{B}_s- B_s$ mixing, but we expect that the left-handed  one dominates, as the right-handed one 
is suppressed by the masses of bottom and strange quarks, i.e.:
\beq
\frac{m_{\tilde b}^2 m_{\tilde s}^2 }{m_{\tilde \psi}^4} \sim \frac{2 \times 10^{-13}}{(m_{\tilde \psi}/ \TeV)^4}
\eeq
The bound on the Wilson coefficient of the first operator is given by  (see Table 1.1 of \Ref{Isidori:2013ez} and \cite{Lenz:2010gu}):
\beq
\frac{g_D^2}{ 4 m_{Z^\prime}^2} s_{\theta_2}^2  c_{\theta_2}^2   t_{\theta_3}^2 < 5.0 \times 10^{-5} \TeV^{-2}
\eeq
Combined with Eq.~(\ref{eq:RKrequirement}), we can derive the bounds on the combination of the mixing angles or the 
factor $g_D^2/m_{Z'}^2$ :
\beq
 s_{\theta_2}c_{\theta_2}t_{\theta_3}  < 0.082 \qquad \Rightarrow \qquad m_{Z'} < g_D \times 5.8 \TeV
\eeq
which is satisfied by our benchmark point ($s_{\theta_2}c_{\theta_2}t_{\theta_3} \simeq 0.041$, $m_{Z'}/g_D \simeq  4.1$~TeV). 

Next, we consider the searches for $tZ$ flavor changing neutral current in the top quark decays at LHC. The relevant interactions are 
\begin{align}
\mathcal{L}_{tZ}&=  g   \frac{m_{\tilde t} m_{\tilde c}}{2 m^2_{\tilde \psi}}   t_{\theta_2} t_{\theta_3} Z_\mu \bar{\tilde t}_R\gamma^\mu \tilde c_R  + H.c. \,,
\label{eq:LtZ}
\end{align} 
which will contribute to the  decay channel $\tilde t\rightarrow Z \tilde c$ with branching ratio estimated as:
\beq
\text{BR}(\tilde t \rightarrow Z \tilde c) \sim  \frac{m_{\tilde t}^2 m_{\tilde c}^2}{2 m^4_{\tilde \psi}}   t^2_{\theta_2} t_{\theta_3}^2
\eeq
 The most stringent bound on such branching ratio comes from ATLAS search at 13 TeV (36 fb$^{-1}$)~\cite{Durglishvili:2017tym}. 
 The results read
\beq
\text{BR}(\tilde t\rightarrow Z \tilde c) < 2.3 \times 10^{-4} \,,
\eeq
which sets a lower bound on our parameters,
\beq
\frac{m_{\tilde \psi} }{ \sqrt{ t_{\theta_2} t_{\theta_3}}} > 102\GeV \,.
\eeq
It is easy to see that such bound is satisfied in the region of parameters consistent with our benchmark point.

Finally, we discuss the constraints from flavor physics in the lepton sector. The gauge boson $Z'$ under $U(1)_{L_\mu-L_\tau}$ 
can contribute to tau lepton decay $\tau \to \mu \nu_\tau \bar{\nu}_\mu$ via 1-loop box diagram~\cite{Altmannshofer:2014cfa}.
It excludes $Z'$ mass lower than 650 GeV for $g_D =1$, and the benchmark point with a 4 TeV $Z'$ is quite safe with this limit.

\section{Conclusion and Summary}
\label{sec:Conclusions}

The recently observed anomalies in the decay of $B$-mesons into $K$-mesons and pair of leptons
could be a possible hint of the violation of lepton-universality, one of the pillars of the SM phenomenology.
In this work, we have provided a possible realization of the idea of explaining these anomalies by the
presence of a new gauge boson $Z'$ of an $U(1)_{L_\mu - L_\tau}$ symmetry, that couple to leptons at tree-level, 
and to quarks via the mixing with vector-like quarks charged under the new gauge symmetry. The mixing of SM 
quarks with vector-like  quarks needs to be induced by the presence of a new Higgs boson, charged under the new 
$U(1)_{L_\mu - L_\tau}$ symmetry.

In this work, we have discussed the possibility that such Higgs boson could be associated with the
recent excess in the di-photon channel observed at the CMS experiment at the 13~TeV LHC run. We have shown
that this is possible and provided the constraints on the quark-mixing parameters that could lead
to such a possibility. An additional, naturally heavy, Higgs boson, contributing to the breakdown of
the $U(1)_{L_\mu - L_\tau}$ symmetry, but not to the quark mixing,  is necessary in order to give the $Z'$ 
gauge boson a 
sufficiently high mass to avoid the LHC constraints. A simultaneous explanation of these observables with the 
Higgs boson excess   observed at LEP may be achieved in a way that is consistent with most observations, although
it is in tension with the rate of the SM-like Higgs boson decaying to di-photons in the gluon fusion production channel
measured by the ATLAS experiment in the 13~TeV LHC run.

\section*{Acknowledgments}
Work at University of Chicago is supported in part by U.S. Department of Energy grant number DE-FG02-13ER41958. 
Work at ANL is supported in part by the U.S. Department of Energy under Contract No. DE-AC02-06CH11357.  
The work of DL was partially performed at the Aspen Center for Physics, which is supported by National Science 
Foundation grant PHY-1607611. We would like to thank Zhen Liu, Emmanuel Stamou, LianTao Wang 
for useful discussions and comments. JL acknowledges support by an Oehme Fellowship.

\appendix

\section{Scalar mixing parameters}
\label{sec:scalarmixing}

In this section, we give the relation between the  five free parameters $\mu_\phi, \mu, \lambda_\phi
, \lambda$, $\lambda_{\phi h}$ and the five physical observables $v_D$, $v$, $m_{\tilde{h}}$,
$m_{\tilde{\phi}}$, $\sin \theta$. The relation is given below,

\begin{align}
v_D&=\left(-\frac{\mu_\phi^2}{\lambda_\phi}\right)^{\frac{1}{2}} , \quad 
v=\left(-\frac{\mu^2}{\lambda}\right)^{\frac{1}{2}}  , \nonumber \\
m_h&=\sqrt{2}\mu = -\sqrt{2}\lambda v ~ , \quad m_\phi=\sqrt{2}\mu_\phi  = -\sqrt{2}\lambda_\phi v_D  , \nonumber \\
m^2_{\tilde h} &= \frac{1}{2}\left(m^2_{\phi}+ m^2_h +\sqrt{\left( m^2_h-m^2_{\phi} \right)^2+ 4\lambda^2_{\phi h} v^2v^2_D}\right)  , \nonumber \\
m^2_{\tilde \phi} &= \frac{1}{2}\left(m^2_{\phi}+ m^2_h -\sqrt{\left( m^2_h-m^2_{\phi} \right)^2+ 4\lambda^2_{\phi h} v^2v^2_D}\right) , \nonumber \\
\tan2\alpha&=\frac{2\lambda_{\phi h} vv_D}{m^2_h-m^2_\phi} .
\end{align}

\section{Limits from CKM matrix}
\label{sec:CKM}

The most general interactions for $U(1)_{\mu-\tau}$ within the quark sector are given by:
\begin{align}
\mathcal{L}_q & = i \bar{q}^i_L \slashed D q_L^i +  i \bar{u}^i_R \slashed D u_R^i +  i \bar{d}^i_R \slashed D d_R^i  
+  i \bar{\psi} \slashed D \psi - m_{\psi}  \bar{\psi} \psi  
- \left( \bar{q}^i_L y^{ij}_{u} \tilde H u_R^j + \bar{q}_L^i y_{d}^{ij} H d_R^{j} + H.c.\right)  \nonumber  \\ 
&+\lambda_1 \bar{q_L^1}\phi^\star \psi_R+\lambda_2 \bar{q_L^2}\phi^\star \psi_R 
+\lambda_3 \bar{q_L^3} \phi \psi_R +H.c. \,. 
\end{align}
The mass matrix to generate CKM matrix in the quark sector are written in the basis of $( q_3, q_2, q_1, \psi)$. 
For up-type and down-type quarks, the mass matrices are 
\begin{align}
M_u=\left(\begin{array}{cccc}
m_t & 0 & 0 &\lambda_3 v_D \\
m_{23}^u & m_c & 0 & \lambda_2 v_D\\
0 & 0 & m_u & \lambda_1 v_D\\
0 & 0 & 0 &m_\psi
\end{array}\right)  , \quad 
M_d=\left(\begin{array}{cccc}
m_b & 0 & 0 &\lambda_3 v_D \\
0 & m_s & 0 & \lambda_2 v_D\\
m_{13}^d & m_{12}^d & m_d & \lambda_1 v_D\\
0 & 0 & 0 &m_\psi
\end{array}\right) \,.
\label{eq:fullMforCKM}
\end{align}
In principle, all the matrix elements except the fourth line can be non-zero. Here we consider the simple matrices 
in Eq.~(\ref{eq:fullMforCKM}) to generate correct CKM matrix, with only three non-zero parameters in the off-diagonal
terms for SM quark mass. To avoid modifying our existing results, we can set $\lambda_1 = 0$. 
Our notation is $M_{\text{diagonal}} = U_L^\dagger M U_R$, thus $q^{\text{flavor}}_{R} 
= U_R q^{\text{mass}}_{R}$ and $q^{\text{flavor}}_{L} = U_L q^{\text{mass}}_{L}$. Therefore, 
we have $\bar{u}^{\text{flavor}}_{L}\gamma^\mu W^{+}_\mu d^{\text{flavor}}_{L} =
\bar{u}^{\text{mass}}_{L}\gamma^\mu W^{+}_\mu V_{\text{CKM}} d^{\text{mass}}_{L} $, which results in the CKM matrix  
$V_{\text{CKM}} = U_L^{u\dagger} U_L^d$. 
The mass of quarks are slightly changed by the small non-zero off-diagonal terms $m_{23}^u$, $m_{13}^d$ and $m_{12}^d$.
The SM quark masses at leading order are
\begin{align}
m_{\tilde t}^2 &= m_t^2  - \frac{v_D^2 \left(m^u_{23} \lambda_2 + m_t \lambda_3 \right)^2}{v_D^2 \lambda_2^2 + m_\psi^2 -m_t^2} 
+ \left( m_{23}^{u} \right)^2 ,   \quad
m_{\tilde c}^2 = m_c^2 \cos^2\theta_2  , \quad m_{\tilde{u}} = m_u  \\
m_{\tilde b}^2 &= m_b^2 \left(1 - \frac{\lambda_3^2}{\lambda_2^2}\right)+ \left( m_{13}^{d} \right)^2  , \quad
m_{\tilde s}^2 = \cot^2\theta_2  \frac{m_{d}^2 }{(m_{12}^{d})^2+m_d^2} m_s^2  , \nonumber \\
m_{\tilde d}^2 & = m_d^2 + (m_{12}^d)^2 \left(1+ \cot^2\theta_2 \frac{m_s^2}{m_d^2 + (m_{12}^d)^2 } \right) \,.
\end{align}
The heavy vector-like quark mass are the same as in Eq.~(\ref{eq:heavyQmass}). The expressions for off-diagonal CKM 
matrix at leading order are given by,
\begin{align}
V_{ub}&= \frac{m_{13}^d}{m_b}  ,  \\
V_{us}&=\frac{m_{12}^dm_s \cot\theta_2}{m_{12}^{d,2} + m_d^2}  ,  \\
V_{cd}&=  - \cot\theta_2 \frac{m_{12}^d  m_s }{(m_{12}^{d})^2 + m_d^2} 
- \cot\theta_2 \frac{m_{13}^d  m_{23}^u }{m_b m_t} \simeq -V_{us},  \\
V_{cb}&= \cot\theta_2 \frac{m_{23}^u }{m_t} , \\ 
V_{ts} &= -\cot\theta_2 \frac{m_{23}^u}{m_t} - \cot\theta_2 \frac{m_{13}^d}{m_b}\frac{m_{12}^dm_s }{(m_{12}^{d})^2 + m_d^2} 
\simeq -V_{cb} ,  \\ 
V_{td} &= \frac{m_{23}^u \cot\theta_2}{m_t}\frac{m_{12}^dm_s \cot\theta_2}{m_{12}^{d,2} + m_d^2} -\frac{m_{13}^{d,\star}}{m_b} \simeq V_{cb} \times V_{us} - V_{ub}^\star .
\end{align}

We require $|V_{us}|\sim |V_{cd}|\simeq 0.225$,  $|V_{cb}|\sim |V_{ts}| \simeq 0.043$, $|V_{ub}|\sim 0.004$
to satisfy the CKM matrix from PDG \cite{Patrignani:2016xqp}. To realize $|V_{td}|\sim 0.008$, we need $m_{13}^d$ 
to be complex number. For our benchmark point $\cot\theta_2 \sim 0.512$, the non-zero off-diagonal SM quark mass
required by CKM matrix are
\begin{align}
m_{12}^d & \simeq 0.01 ~m_{\tilde d}  , \quad
|m_{13}^d|  \simeq 0.004 ~m_{\tilde b}, \quad
m_{23}^u \simeq 0.084 ~m_{\tilde t}.
\end{align}
It is clear that they are sufficiently small comparing with the diagonal masses. Thus we successfully
demonstrate that the generation of CKM matrix does not affect the new physics phenomenology 
we have discussed previously.

\bibliography{referencelist}

\providecommand{\href}[2]{#2}\begingroup\raggedright\begin{thebibliography}{10}

\bibitem{CMS:2017yta}
{\bf CMS} Collaboration, {\it {Search for new resonances in the diphoton final
  state in the mass range between 70 and 110 GeV in pp collisions at
  $\sqrt{s}=$ 8 and 13 TeV}},  Tech. Rep. CMS-PAS-HIG-17-013, CERN, Geneva,
  2017.

\bibitem{CMS-PAS-HIG-14-037}
{\bf CMS Collaboration} Collaboration, {\it {Search for new resonances in the
  diphoton final state in the mass range between 80 and 110 GeV in pp
  collisions at $\sqrt{s}=8$ TeV}},  Tech. Rep. CMS-PAS-HIG-14-037, CERN,
  Geneva, 2015.

\bibitem{Aad:2014ioa}
{\bf ATLAS} Collaboration, G.~Aad et~al., {\it {Search for Scalar Diphoton
  Resonances in the Mass Range $65-600$ GeV with the ATLAS Detector in $pp$
  Collision Data at $\sqrt{s}$ = 8 $TeV$}},  {\em Phys. Rev. Lett.} {\bf 113}
  (2014), no.~17 171801, [\href{http://arxiv.org/abs/1407.6583}{{\tt
  arXiv:1407.6583}}].

\bibitem{Barate:2003sz}
{\bf OPAL, DELPHI, LEP Working Group for Higgs boson searches, ALEPH, L3}
  Collaboration, R.~Barate et~al., {\it {Search for the standard model Higgs
  boson at LEP}},  {\em Phys. Lett.} {\bf B565} (2003) 61--75,
  [\href{http://arxiv.org/abs/hep-ex/0306033}{{\tt hep-ex/0306033}}].

\bibitem{Cao:2016uwt}
J.~Cao, X.~Guo, Y.~He, P.~Wu, and Y.~Zhang, {\it {Diphoton signal of the light
  Higgs boson in natural NMSSM}},  {\em Phys. Rev.} {\bf D95} (2017), no.~11
  116001, [\href{http://arxiv.org/abs/1612.08522}{{\tt arXiv:1612.08522}}].

\bibitem{Mariotti:2017vtv}
A.~Mariotti, D.~Redigolo, F.~Sala, and K.~Tobioka, {\it {New LHC bound on
  low-mass diphoton resonances}},  \href{http://arxiv.org/abs/1710.01743}{{\tt
  arXiv:1710.01743}}.

\bibitem{Crivellin:2017upt}
A.~Crivellin, J.~Heeck, and D.~Müller, {\it {Large $h\to b s$ in generic
  two-Higgs-doublet models}},  {\em Phys. Rev.} {\bf D97} (2018), no.~3 035008,
  [\href{http://arxiv.org/abs/1710.04663}{{\tt arXiv:1710.04663}}].

\bibitem{Fox:2017uwr}
P.~J. Fox and N.~Weiner, {\it {Light Signals from a Lighter Higgs}},
  \href{http://arxiv.org/abs/1710.07649}{{\tt arXiv:1710.07649}}.

\bibitem{Haisch:2017gql}
U.~Haisch and A.~Malinauskas, {\it {Let there be light from a second light
  Higgs doublet}},  {\em JHEP} {\bf 03} (2018) 135,
  [\href{http://arxiv.org/abs/1712.06599}{{\tt arXiv:1712.06599}}].

\bibitem{Vega:2018ddp}
R.~Vega, R.~Vega-Morales, and K.~Xie, {\it {Light (and darkness) from a light
  hidden Higgs}},  \href{http://arxiv.org/abs/1805.01970}{{\tt
  arXiv:1805.01970}}.

\bibitem{Aaij:2014ora}
{\bf LHCb} Collaboration, R.~Aaij et~al., {\it {Test of lepton universality
  using $B^{+}\rightarrow K^{+}\ell^{+}\ell^{-}$ decays}},  {\em Phys. Rev.
  Lett.} {\bf 113} (2014) 151601, [\href{http://arxiv.org/abs/1406.6482}{{\tt
  arXiv:1406.6482}}].

\bibitem{Aaij:2017vbb}
{\bf LHCb} Collaboration, R.~Aaij et~al., {\it {Test of lepton universality
  with $B^{0} \rightarrow K^{*0}\ell^{+}\ell^{-}$ decays}},  {\em JHEP} {\bf
  08} (2017) 055, [\href{http://arxiv.org/abs/1705.05802}{{\tt
  arXiv:1705.05802}}].

\bibitem{Altmannshofer:2014cfa}
W.~Altmannshofer, S.~Gori, M.~Pospelov, and I.~Yavin, {\it {Quark flavor
  transitions in $L_\mu-L_\tau$ models}},  {\em Phys. Rev.} {\bf D89} (2014)
  095033, [\href{http://arxiv.org/abs/1403.1269}{{\tt arXiv:1403.1269}}].

\bibitem{Crivellin:2015mga}
A.~Crivellin, G.~D'Ambrosio, and J.~Heeck, {\it {Explaining
  $h\to\mu^\pm\tau^\mp$, $B\to K^* \mu^+\mu^-$ and $B\to K \mu^+\mu^-/B\to K
  e^+e^-$ in a two-Higgs-doublet model with gauged $L_\mu-L_\tau$}},  {\em
  Phys. Rev. Lett.} {\bf 114} (2015) 151801,
  [\href{http://arxiv.org/abs/1501.00993}{{\tt arXiv:1501.00993}}].

\bibitem{Altmannshofer:2015mqa}
W.~Altmannshofer and I.~Yavin, {\it {Predictions for lepton flavor universality
  violation in rare B decays in models with gauged $L_\mu - L_\tau$}},  {\em
  Phys. Rev.} {\bf D92} (2015), no.~7 075022,
  [\href{http://arxiv.org/abs/1508.07009}{{\tt arXiv:1508.07009}}].

\bibitem{Altmannshofer:2016oaq}
W.~Altmannshofer, M.~Carena, and A.~Crivellin, {\it {$L_\mu - L_\tau$ theory of
  Higgs flavor violation and $(g-2)_\mu$}},  {\em Phys. Rev.} {\bf D94} (2016),
  no.~9 095026, [\href{http://arxiv.org/abs/1604.08221}{{\tt
  arXiv:1604.08221}}].

\bibitem{Altmannshofer:2016jzy}
W.~Altmannshofer, S.~Gori, S.~Profumo, and F.~S. Queiroz, {\it {Explaining dark
  matter and B decay anomalies with an $L_\mu - L_\tau$ model}},  {\em JHEP}
  {\bf 12} (2016) 106, [\href{http://arxiv.org/abs/1609.04026}{{\tt
  arXiv:1609.04026}}].

\bibitem{Alonso:2017uky}
R.~Alonso, P.~Cox, C.~Han, and T.~T. Yanagida, {\it {Flavoured $B-L$ local
  symmetry and anomalous rare $B$ decays}},  {\em Phys. Lett.} {\bf B774}
  (2017) 643--648, [\href{http://arxiv.org/abs/1705.03858}{{\tt
  arXiv:1705.03858}}].

\bibitem{Bonilla:2017lsq}
C.~Bonilla, T.~Modak, R.~Srivastava, and J.~W.~F. Valle, {\it
  {$U(1)_{B_3-3L_\mu}$ gauge symmetry as the simplest description of $b\to s$
  anomalies}},  \href{http://arxiv.org/abs/1705.00915}{{\tt arXiv:1705.00915}}.

\bibitem{Nomura:2018vfz}
T.~Nomura and H.~Okada, {\it {A Zee-Babu type model with $U(1)_{L_\mu -
  L_\tau}$ gauge symmetry}},  \href{http://arxiv.org/abs/1803.04795}{{\tt
  arXiv:1803.04795}}.

\bibitem{Chen:2017usq}
C.-H. Chen and T.~Nomura, {\it {Penguin $b \to s\ell'^+ \ell'^-$ and $B$-meson
  anomalies in a gauged ${L_\mu -L_\tau}$}},  {\em Phys. Lett.} {\bf B777}
  (2018) 420--427, [\href{http://arxiv.org/abs/1707.03249}{{\tt
  arXiv:1707.03249}}].

\bibitem{Ko:2017yrd}
P.~Ko, T.~Nomura, and H.~Okada, {\it {Explaining $B\to K^{(*)}\ell^+ \ell^-$
  anomaly by radiatively induced coupling in $U(1)_{\mu-\tau}$ gauge
  symmetry}},  {\em Phys. Rev.} {\bf D95} (2017), no.~11 111701,
  [\href{http://arxiv.org/abs/1702.02699}{{\tt arXiv:1702.02699}}].

\bibitem{Crivellin:2015lwa}
A.~Crivellin, G.~D'Ambrosio, and J.~Heeck, {\it {Addressing the LHC flavor
  anomalies with horizontal gauge symmetries}},  {\em Phys. Rev.} {\bf D91}
  (2015), no.~7 075006, [\href{http://arxiv.org/abs/1503.03477}{{\tt
  arXiv:1503.03477}}].

\bibitem{Celis:2015ara}
A.~Celis, J.~Fuentes-Martin, M.~Jung, and H.~Serodio, {\it {Family nonuniversal
  $Z'$ models with protected flavor-changing interactions}},  {\em Phys. Rev.}
  {\bf D92} (2015), no.~1 015007, [\href{http://arxiv.org/abs/1505.03079}{{\tt
  arXiv:1505.03079}}].

\bibitem{Falkowski:2015zwa}
A.~Falkowski, M.~Nardecchia, and R.~Ziegler, {\it {Lepton Flavor
  Non-Universality in B-meson Decays from a U(2) Flavor Model}},  {\em JHEP}
  {\bf 11} (2015) 173, [\href{http://arxiv.org/abs/1509.01249}{{\tt
  arXiv:1509.01249}}].

\bibitem{Alok:2017sui}
A.~K. Alok, B.~Bhattacharya, A.~Datta, D.~Kumar, J.~Kumar, and D.~London, {\it
  {New Physics in $b \to s \mu^+ \mu^-$ after the Measurement of $R_{K^*}$}},
  {\em Phys. Rev.} {\bf D96} (2017), no.~9 095009,
  [\href{http://arxiv.org/abs/1704.07397}{{\tt arXiv:1704.07397}}].

\bibitem{PhysRevD.44.2118}
X.-G. He, G.~C. Joshi, H.~Lew, and R.~R. Volkas, {\it Simplest
  ${Z}^{\ensuremath{'}}$ model},  {\em Phys. Rev. D} {\bf 44} (Oct, 1991)
  2118--2132.

\bibitem{Baek:2001kca}
S.~Baek, N.~G. Deshpande, X.~G. He, and P.~Ko, {\it {Muon anomalous g-2 and
  gauged $L_\mu-L_\tau$ models}},  {\em Phys. Rev.} {\bf D64} (2001) 055006,
  [\href{http://arxiv.org/abs/hep-ph/0104141}{{\tt hep-ph/0104141}}].

\bibitem{Foot:1994vd}
R.~Foot, X.~G. He, H.~Lew, and R.~R. Volkas, {\it {Model for a light Z-prime
  boson}},  {\em Phys. Rev.} {\bf D50} (1994) 4571--4580,
  [\href{http://arxiv.org/abs/hep-ph/9401250}{{\tt hep-ph/9401250}}].

\bibitem{Salvioni:2009jp}
E.~Salvioni, A.~Strumia, G.~Villadoro, and F.~Zwirner, {\it {Non-universal
  minimal Z' models: present bounds and early LHC reach}},  {\em JHEP} {\bf 03}
  (2010) 010, [\href{http://arxiv.org/abs/0911.1450}{{\tt arXiv:0911.1450}}].

\bibitem{Heeck:2011wj}
J.~Heeck and W.~Rodejohann, {\it {Gauged $L_\mu-L_\tau$ Symmetry at the
  Electroweak Scale}},  {\em Phys. Rev.} {\bf D84} (2011) 075007,
  [\href{http://arxiv.org/abs/1107.5238}{{\tt arXiv:1107.5238}}].

\bibitem{PhysRevD.85.115016}
W.-Z. Feng, P.~Nath, and G.~Peim, {\it Cosmic coincidence and asymmetric dark
  matter in a stueckelberg extension},  {\em Phys. Rev. D} {\bf 85} (Jun, 2012)
  115016.

\bibitem{Harigaya:2013twa}
K.~Harigaya, T.~Igari, M.~M. Nojiri, M.~Takeuchi, and K.~Tobe, {\it {Muon g-2
  and LHC phenomenology in the $L_\mu-L_\tau$ gauge symmetric model}},  {\em
  JHEP} {\bf 03} (2014) 105, [\href{http://arxiv.org/abs/1311.0870}{{\tt
  arXiv:1311.0870}}].

\bibitem{Ma:2001md}
E.~Ma, D.~P. Roy, and S.~Roy, {\it {Gauged $L_\mu - L_\tau$ with large muon
  anomalous magnetic moment and the bimaximal mixing of neutrinos}},  {\em
  Phys. Lett.} {\bf B525} (2002) 101--106,
  [\href{http://arxiv.org/abs/hep-ph/0110146}{{\tt hep-ph/0110146}}].

\bibitem{Ibe:2016dir}
M.~Ibe, W.~Nakano, and M.~Suzuki, {\it {Constraints on $L_\mu-L_\tau$ gauge
  interactions from rare kaon decay}},  {\em Phys. Rev.} {\bf D95} (2017),
  no.~5 055022, [\href{http://arxiv.org/abs/1611.08460}{{\tt
  arXiv:1611.08460}}].

\bibitem{Biswas:2016yan}
A.~Biswas, S.~Choubey, and S.~Khan, {\it {Neutrino Mass, Dark Matter and
  Anomalous Magnetic Moment of Muon in a $U(1)_{L_{\mu}-L_{\tau}}$ Model}},
  {\em JHEP} {\bf 09} (2016) 147, [\href{http://arxiv.org/abs/1608.04194}{{\tt
  arXiv:1608.04194}}].

\bibitem{Elahi:2015vzh}
F.~Elahi and A.~Martin, {\it {Constraints on $L_\mu - L_\tau$ interactions at
  the LHC and beyond}},  {\em Phys. Rev.} {\bf D93} (2016), no.~1 015022,
  [\href{http://arxiv.org/abs/1511.04107}{{\tt arXiv:1511.04107}}].

\bibitem{Gninenko:2018tlp}
S.~N. Gninenko and N.~V. Krasnikov, {\it {Probing the muon $g_\mu-2$ anomaly,
  $L_{\mu} - L_{\tau}$ gauge boson and Dark Matter in dark photon
  experiments}},  \href{http://arxiv.org/abs/1801.10448}{{\tt
  arXiv:1801.10448}}.

\bibitem{Cao:2017sju}
J.~Cao, L.~Feng, X.~Guo, L.~Shang, F.~Wang, P.~Wu, and L.~Zu, {\it {Explaining
  the DAMPE data with scalar dark matter and gauged $U(1)_{L_e-L_\mu }$
  interaction}},  {\em Eur. Phys. J.} {\bf C78} (2018), no.~3 198,
  [\href{http://arxiv.org/abs/1712.01244}{{\tt arXiv:1712.01244}}].

\bibitem{Biswas:2017ait}
A.~Biswas, S.~Choubey, L.~Covi, and S.~Khan, {\it {Explaining the 3.5 keV X-ray
  Line in a ${L_{\mu}-L_{\tau}}$ Extension of the Inert Doublet Model}},  {\em
  JCAP} {\bf 1802} (2018), no.~02 002,
  [\href{http://arxiv.org/abs/1711.00553}{{\tt arXiv:1711.00553}}].

\bibitem{Baek:2017sew}
S.~Baek, {\it {Dark matter contribution to $b\to s \mu^+ \mu^-$ anomaly in
  local $U(1)_{L_\mu-L_\tau}$ model}},  {\em Phys. Lett.} {\bf B781} (2018)
  376--382, [\href{http://arxiv.org/abs/1707.04573}{{\tt arXiv:1707.04573}}].

\bibitem{Elahi:2017ppe}
F.~Elahi and A.~Martin, {\it {Using the modified matrix element method to
  constrain $L_\mu - L_\tau$ interactions}},  {\em Phys. Rev.} {\bf D96}
  (2017), no.~1 015021, [\href{http://arxiv.org/abs/1705.02563}{{\tt
  arXiv:1705.02563}}].

\bibitem{Asai:2017ryy}
K.~Asai, K.~Hamaguchi, and N.~Nagata, {\it {Predictions for the neutrino
  parameters in the minimal gauged U(1)$_{L_\mu-L_\tau}$ model}},  {\em Eur.
  Phys. J.} {\bf C77} (2017), no.~11 763,
  [\href{http://arxiv.org/abs/1705.00419}{{\tt arXiv:1705.00419}}].

\bibitem{Kaneta:2016uyt}
Y.~Kaneta and T.~Shimomura, {\it {On the possibility of a search for the $L_\mu
  - L_\tau$ gauge boson at Belle-II and neutrino beam experiments}},  {\em
  PTEP} {\bf 2017} (2017), no.~5 053B04,
  [\href{http://arxiv.org/abs/1701.00156}{{\tt arXiv:1701.00156}}].

\bibitem{Biswas:2016yjr}
A.~Biswas, S.~Choubey, and S.~Khan, {\it {FIMP and Muon ($g-2$) in a
  U$(1)_{L_{\mu}-L_{\tau}}$ Model}},  {\em JHEP} {\bf 02} (2017) 123,
  [\href{http://arxiv.org/abs/1612.03067}{{\tt arXiv:1612.03067}}].

\bibitem{Tang:2017gkz}
Y.~Tang and Y.-L. Wu, {\it {Flavor non-universal gauge interactions and
  anomalies in B-meson decays}},  {\em Chin. Phys.} {\bf C42} (2018), no.~3
  033104, [\href{http://arxiv.org/abs/1705.05643}{{\tt arXiv:1705.05643}}].

\bibitem{Ko:2017quv}
P.~Ko, T.~Nomura, and H.~Okada, {\it {A flavor dependent gauge symmetry,
  Predictive radiative seesaw and LHCb anomalies}},  {\em Phys. Lett.} {\bf
  B772} (2017) 547--552, [\href{http://arxiv.org/abs/1701.05788}{{\tt
  arXiv:1701.05788}}].

\bibitem{Arcadi:2018tly}
G.~Arcadi, T.~Hugle, and F.~S. Queiroz, {\it {The Dark $L_\mu - L_\tau$ Rises
  via Kinetic Mixing}},  \href{http://arxiv.org/abs/1803.05723}{{\tt
  arXiv:1803.05723}}.

\bibitem{Kamada:2018zxi}
A.~Kamada, K.~Kaneta, K.~Yanagi, and H.-B. Yu, {\it {Self-interacting dark
  matter and muon $g-2$ in a gauged U$(1)_{L_{\mu} - L_{\tau}}$ model}},
  \href{http://arxiv.org/abs/1805.00651}{{\tt arXiv:1805.00651}}.

\bibitem{Biswas:2018yus}
A.~Biswas, S.~Choubey, and S.~Khan, {\it {Inverse seesaw and dark matter in a
  gauged ${\rm B-L}$ extension with flavour symmetry}},
  \href{http://arxiv.org/abs/1805.00568}{{\tt arXiv:1805.00568}}.

\bibitem{Xing:2015fdg}
Z.-z. Xing and Z.-h. Zhao, {\it {A review of $\mu$ - $\tau$ flavor symmetry in
  neutrino physics}},  {\em Rept. Prog. Phys.} {\bf 79} (2016), no.~7 076201,
  [\href{http://arxiv.org/abs/1512.04207}{{\tt arXiv:1512.04207}}].

\bibitem{deFlorian:2016spz}
{\bf LHC Higgs Cross Section Working Group} Collaboration, D.~de~Florian
  et~al., {\it {Handbook of LHC Higgs Cross Sections: 4. Deciphering the Nature
  of the Higgs Sector}},  \href{http://arxiv.org/abs/1610.07922}{{\tt
  arXiv:1610.07922}}.

\bibitem{Heinemeyer:2013tqa}
{\bf LHC Higgs Cross Section Working Group} Collaboration, J.~R. Andersen
  et~al., {\it {Handbook of LHC Higgs Cross Sections: 3. Higgs Properties}},
  \href{http://arxiv.org/abs/1307.1347}{{\tt arXiv:1307.1347}}.

\bibitem{CMS-PAS-HIG-17-031}
{\bf CMS Collaboration} Collaboration, {\it {Combined measurements of the Higgs
  boson's couplings at $\sqrt{s}=13$ TeV}},  Tech. Rep. CMS-PAS-HIG-17-031,
  CERN, Geneva, 2018.

\bibitem{ATLAS-CONF-2017-047}
{\bf ATLAS Collaboration} Collaboration, {\it {Combined measurements of Higgs
  boson production and decay in the $H\rightarrow ZZ^* \rightarrow 4\ell$ and
  $H\rightarrow\gamma\gamma$ channels using $\sqrt{s}=$ 13 TeV pp collision
  data collected with the ATLAS experiment}},  Tech. Rep. ATLAS-CONF-2017-047,
  CERN, Geneva, Jul, 2017.

\bibitem{Sirunyan:2018ouh}
{\bf CMS} Collaboration, A.~M. Sirunyan et~al., {\it {Measurements of Higgs
  boson properties in the diphoton decay channel in proton-proton collisions at
  $\sqrt{s} =$ 13 TeV}},  \href{http://arxiv.org/abs/1804.02716}{{\tt
  arXiv:1804.02716}}.

\bibitem{Aaboud:2018xdt}
{\bf ATLAS} Collaboration, M.~Aaboud et~al., {\it {Measurements of Higgs boson
  properties in the diphoton decay channel with 36 fb$^{-1}$ of $pp$ collision
  data at $\sqrt{s} = 13$ TeV with the ATLAS detector}},
  \href{http://arxiv.org/abs/1802.04146}{{\tt arXiv:1802.04146}}.

\bibitem{Aad:2014eha}
{\bf ATLAS} Collaboration, G.~Aad et~al., {\it {Measurement of Higgs boson
  production in the diphoton decay channel in pp collisions at center-of-mass
  energies of 7 and 8 TeV with the ATLAS detector}},  {\em Phys. Rev.} {\bf
  D90} (2014), no.~11 112015, [\href{http://arxiv.org/abs/1408.7084}{{\tt
  arXiv:1408.7084}}].

\bibitem{Altmannshofer:2013foa}
W.~Altmannshofer and D.~M. Straub, {\it {New Physics in $B \to K^*\mu\mu$}},
  {\em Eur. Phys. J.} {\bf C73} (2013) 2646,
  [\href{http://arxiv.org/abs/1308.1501}{{\tt arXiv:1308.1501}}].

\bibitem{Altmannshofer:2014rta}
W.~Altmannshofer and D.~M. Straub, {\it {New physics in $b\rightarrow s$
  transitions after LHC run 1}},  {\em Eur. Phys. J.} {\bf C75} (2015), no.~8
  382, [\href{http://arxiv.org/abs/1411.3161}{{\tt arXiv:1411.3161}}].

\bibitem{Altmannshofer:2015sma}
W.~Altmannshofer and D.~M. Straub, {\it {Implications of $b\to s$
  measurements}},  in {\em {Proceedings, 50th Rencontres de Moriond Electroweak
  Interactions and Unified Theories: La Thuile, Italy, March 14-21, 2015}},
  pp.~333--338, 2015.
\newblock \href{http://arxiv.org/abs/1503.06199}{{\tt arXiv:1503.06199}}.

\bibitem{Greljo:2015mma}
A.~Greljo, G.~Isidori, and D.~Marzocca, {\it {On the breaking of Lepton Flavor
  Universality in B decays}},  {\em JHEP} {\bf 07} (2015) 142,
  [\href{http://arxiv.org/abs/1506.01705}{{\tt arXiv:1506.01705}}].

\bibitem{Descotes-Genon:2015uva}
S.~Descotes-Genon, L.~Hofer, J.~Matias, and J.~Virto, {\it {Global analysis of
  $b\to s\ell\ell$ anomalies}},  {\em JHEP} {\bf 06} (2016) 092,
  [\href{http://arxiv.org/abs/1510.04239}{{\tt arXiv:1510.04239}}].

\bibitem{Hurth:2016fbr}
T.~Hurth, F.~Mahmoudi, and S.~Neshatpour, {\it {On the anomalies in the latest
  LHCb data}},  {\em Nucl. Phys.} {\bf B909} (2016) 737--777,
  [\href{http://arxiv.org/abs/1603.00865}{{\tt arXiv:1603.00865}}].

\bibitem{Altmannshofer:2017yso}
W.~Altmannshofer, P.~Stangl, and D.~M. Straub, {\it {Interpreting Hints for
  Lepton Flavor Universality Violation}},  {\em Phys. Rev.} {\bf D96} (2017),
  no.~5 055008, [\href{http://arxiv.org/abs/1704.05435}{{\tt
  arXiv:1704.05435}}].

\bibitem{Capdevila:2017bsm}
B.~Capdevila, A.~Crivellin, S.~Descotes-Genon, J.~Matias, and J.~Virto, {\it
  {Patterns of New Physics in $b\to s\ell^+\ell^-$ transitions in the light of
  recent data}},  {\em JHEP} {\bf 01} (2018) 093,
  [\href{http://arxiv.org/abs/1704.05340}{{\tt arXiv:1704.05340}}].

\bibitem{Altmannshofer:2017fio}
W.~Altmannshofer, C.~Niehoff, P.~Stangl, and D.~M. Straub, {\it {Status of the
  $B\rightarrow K^*\mu ^+\mu ^-$ anomaly after Moriond 2017}},  {\em Eur. Phys.
  J.} {\bf C77} (2017), no.~6 377, [\href{http://arxiv.org/abs/1703.09189}{{\tt
  arXiv:1703.09189}}].

\bibitem{Wehle:2016yoi}
{\bf Belle} Collaboration, S.~Wehle et~al., {\it {Lepton-Flavor-Dependent
  Angular Analysis of $B\to K^\ast \ell^+\ell^-$}},  {\em Phys. Rev. Lett.}
  {\bf 118} (2017), no.~11 111801, [\href{http://arxiv.org/abs/1612.05014}{{\tt
  arXiv:1612.05014}}].

\bibitem{Khachatryan:2016zqb}
{\bf CMS} Collaboration, V.~Khachatryan et~al., {\it {Search for narrow
  resonances in dilepton mass spectra in proton-proton collisions at $\sqrt{s}$
  = 13 TeV and combination with 8 TeV data}},  {\em Phys. Lett.} {\bf B768}
  (2017) 57--80, [\href{http://arxiv.org/abs/1609.05391}{{\tt
  arXiv:1609.05391}}].

\bibitem{Aaboud:2016cth}
{\bf ATLAS} Collaboration, M.~Aaboud et~al., {\it {Search for high-mass new
  phenomena in the dilepton final state using proton-proton collisions at
  $\sqrt{s}=13$ TeV with the ATLAS detector}},  {\em Phys. Lett.} {\bf B761}
  (2016) 372--392, [\href{http://arxiv.org/abs/1607.03669}{{\tt
  arXiv:1607.03669}}].

\bibitem{Aaboud:2017buh}
{\bf ATLAS} Collaboration, M.~Aaboud et~al., {\it {Search for new high-mass
  phenomena in the dilepton final state using 36 fb$^{-1}$ of proton-proton
  collision data at $ \sqrt{s}=13 $ TeV with the ATLAS detector}},  {\em JHEP}
  {\bf 10} (2017) 182, [\href{http://arxiv.org/abs/1707.02424}{{\tt
  arXiv:1707.02424}}].

\bibitem{Aad:2014cka}
{\bf ATLAS} Collaboration, G.~Aad et~al., {\it {Search for high-mass dilepton
  resonances in pp collisions at $\sqrt{s}=8$ TeV with the ATLAS detector}},
  {\em Phys. Rev.} {\bf D90} (2014), no.~5 052005,
  [\href{http://arxiv.org/abs/1405.4123}{{\tt arXiv:1405.4123}}].

\bibitem{Aad:2012hf}
{\bf ATLAS} Collaboration, G.~Aad et~al., {\it {Search for high-mass resonances
  decaying to dilepton final states in pp collisions at s**(1/2) = 7-TeV with
  the ATLAS detector}},  {\em JHEP} {\bf 11} (2012) 138,
  [\href{http://arxiv.org/abs/1209.2535}{{\tt arXiv:1209.2535}}].

\bibitem{CMS:2017xrr}
{\bf CMS} Collaboration, {\it {Searches for dijet resonances in pp collisions
  at $\sqrt{s}=13~\mathrm{TeV}$ using data collected in 2016.}},  Tech. Rep.
  CMS-PAS-EXO-16-056, CERN, Geneva, 2017.

\bibitem{Altmannshofer:2014pba}
W.~Altmannshofer, S.~Gori, M.~Pospelov, and I.~Yavin, {\it {Neutrino Trident
  Production: A Powerful Probe of New Physics with Neutrino Beams}},  {\em
  Phys. Rev. Lett.} {\bf 113} (2014) 091801,
  [\href{http://arxiv.org/abs/1406.2332}{{\tt arXiv:1406.2332}}].

\bibitem{Geiregat:1990gz}
{\bf CHARM-II} Collaboration, D.~Geiregat et~al., {\it {First observation of
  neutrino trident production}},  {\em Phys. Lett.} {\bf B245} (1990) 271--275.

\bibitem{Mishra:1991bv}
{\bf CCFR} Collaboration, S.~R. Mishra et~al., {\it {Neutrino tridents and W Z
  interference}},  {\em Phys. Rev. Lett.} {\bf 66} (1991) 3117--3120.

\bibitem{Adams:1998yf}
{\bf NuTeV} Collaboration, T.~Adams et~al., {\it {Neutrino trident production
  from NuTeV}},  in {\em {High-energy physics. Proceedings, 29th International
  Conference, ICHEP'98, Vancouver, Canada, July 23-29, 1998. Vol. 1, 2}},
  pp.~631--634, 1998.
\newblock \href{http://arxiv.org/abs/hep-ex/9811012}{{\tt hep-ex/9811012}}.

\bibitem{Fuyuto:2015gmk}
K.~Fuyuto, W.-S. Hou, and M.~Kohda, {\it {$Z'$ induced FCNC decays of top,
  beauty, and strange quarks}},  {\em Phys. Rev.} {\bf D93} (2016), no.~5
  054021, [\href{http://arxiv.org/abs/1512.09026}{{\tt arXiv:1512.09026}}].

\bibitem{Liu:2017xmc}
D.~Liu, J.~Liu, C.~E.~M. Wagner, and X.-P. Wang, {\it {Bottom-quark
  Forward-Backward Asymmetry, Dark Matter and the LHC}},  {\em Phys. Rev.} {\bf
  D97} (2018), no.~5 055021, [\href{http://arxiv.org/abs/1712.05802}{{\tt
  arXiv:1712.05802}}].

\bibitem{Aaboud:2018lpl}
{\bf ATLAS} Collaboration, M.~Aaboud et~al., {\it {Search for
  R-parity-violating supersymmetric particles in multi-jet final states
  produced in $p$-$p$ collisions at $\sqrt{s} =13$ TeV using the ATLAS detector
  at the LHC}},  \href{http://arxiv.org/abs/1804.03568}{{\tt
  arXiv:1804.03568}}.

\bibitem{Aad:2015lea}
{\bf ATLAS} Collaboration, G.~Aad et~al., {\it {Search for massive
  supersymmetric particles decaying to many jets using the ATLAS detector in
  $pp$ collisions at $\sqrt{s} = 8$ TeV}},  {\em Phys. Rev.} {\bf D91} (2015),
  no.~11 112016, [\href{http://arxiv.org/abs/1502.05686}{{\tt
  arXiv:1502.05686}}]. [Erratum: Phys. Rev.D93,no.3,039901(2016)].

\bibitem{Chatrchyan:2013gia}
{\bf CMS} Collaboration, S.~Chatrchyan et~al., {\it {Searches for light- and
  heavy-flavour three-jet resonances in pp collisions at $\sqrt{s} = 8$ TeV}},
  {\em Phys. Lett.} {\bf B730} (2014) 193--214,
  [\href{http://arxiv.org/abs/1311.1799}{{\tt arXiv:1311.1799}}].

\bibitem{ElHedri:2017nny}
S.~El~Hedri, A.~Kaminska, M.~de~Vries, and J.~Zurita, {\it {Simplified
  Phenomenology for Colored Dark Sectors}},  {\em JHEP} {\bf 04} (2017) 118,
  [\href{http://arxiv.org/abs/1703.00452}{{\tt arXiv:1703.00452}}].

\bibitem{Czakon:2013tha}
M.~Czakon, M.~L. Mangano, A.~Mitov, and J.~Rojo, {\it {Constraints on the gluon
  PDF from top quark pair production at hadron colliders}},  {\em JHEP} {\bf
  07} (2013) 167, [\href{http://arxiv.org/abs/1303.7215}{{\tt
  arXiv:1303.7215}}].

\bibitem{Barbieri:1993av}
R.~Barbieri and G.~F. Giudice, {\it {$b \to s \gamma$ decay and
  supersymmetry}},  {\em Phys. Lett.} {\bf B309} (1993) 86--90,
  [\href{http://arxiv.org/abs/hep-ph/9303270}{{\tt hep-ph/9303270}}].

\bibitem{Patrignani:2016xqp}
{\bf Particle Data Group} Collaboration, C.~Patrignani et~al., {\it {Review of
  Particle Physics}},  {\em Chin. Phys.} {\bf C40} (2016), no.~10 100001.

\bibitem{Buras:2012jb}
A.~J. Buras, F.~De~Fazio, and J.~Girrbach, {\it {The Anatomy of Z' and Z with
  Flavour Changing Neutral Currents in the Flavour Precision Era}},  {\em JHEP}
  {\bf 02} (2013) 116, [\href{http://arxiv.org/abs/1211.1896}{{\tt
  arXiv:1211.1896}}].

\bibitem{Isidori:2013ez}
G.~Isidori, {\it {Flavor physics and CP violation}},  in {\em {Proceedings,
  2012 European School of High-Energy Physics (ESHEP 2012): La Pommeraye,
  Anjou, France, June 06-19, 2012}}, pp.~69--105, 2014.
\newblock \href{http://arxiv.org/abs/1302.0661}{{\tt arXiv:1302.0661}}.

\bibitem{Lenz:2010gu}
A.~Lenz, U.~Nierste, J.~Charles, S.~Descotes-Genon, A.~Jantsch, C.~Kaufhold,
  H.~Lacker, S.~Monteil, V.~Niess, and S.~T'Jampens, {\it {Anatomy of New
  Physics in $B - \bar{B}$ mixing}},  {\em Phys. Rev.} {\bf D83} (2011) 036004,
  [\href{http://arxiv.org/abs/1008.1593}{{\tt arXiv:1008.1593}}].

\bibitem{Durglishvili:2017tym}
{\bf ATLAS} Collaboration, A.~Durglishvili, {\it {Search for $tZ$ Flavour
  Changing Neutral Currents in top-quark decays with the ATLAS detector}},
  \href{http://arxiv.org/abs/1712.09802}{{\tt arXiv:1712.09802}}.

\end{thebibliography}\endgroup

\bibliographystyle{JHEP}   

\end{document}